\documentclass[9pt,a4paper,twoside]{tau-class/tau}
\usepackage[english]{babel}
\usepackage{amsfonts}
\usepackage{amsmath}
\usepackage{hyperref}
\usepackage{bm}
\usepackage{lineno}
\usepackage{cite}

\title{Nonequilibrium physics of brain dynamics}

\author[a,b]{Ramón Nartallo-Kaluarachchi}
\author[b,c,d]{Morten L. Kringelbach}
\author[e,f,g]{Gustavo Deco}
\author[a]{Renaud Lambiotte}
\author[a]{Alain Goriely}

\affil[a]{Mathematical Institute, University of Oxford, Woodstock Road, Oxford, OX2 6GG, United Kingdom}
\affil[b]{Centre for Eudaimonia and Human Flourishing, University of Oxford, 7 Stoke Pl, Oxford, OX3 9BX, United Kingdom}
\affil[c]{Center
for Music in the Brain, Aarhus University, \& The
Royal Academy of Music, Aarhus/Aalborg, Denmark}
\affil[d]{Department of Psychiatry, University of Oxford, Oxford, OX3 7JX United Kingdom}
\affil[e]{Centre for Brain and Cognition, Computational Neuroscience Group, Universitat Pompeu Fabra, 08018
Barcelona, Spain}
\affil[f]{Department of Information and
Communication Technologies, Universitat Pompeu
Fabra, 08018 Barcelona, Spain}
\affil[g]{Institucio Catalana de la Recerca i Estudis Avancats (ICREA), 08010 Barcelona, Spain}
\professor{Correspondence to nartallokalu@maths.ox.ac.uk, goriely@maths.ox.ac.uk}

\leadauthor{Nartallo-Kaluarachchi et al.}

\begin{abstract}    
Information processing in the brain is coordinated by the dynamic activity of neurons and neural populations at a range of spatiotemporal scales. These dynamics, captured in the form of electrophysiological
recordings and neuroimaging, show evidence of time-irreversibility and broken detailed balance suggesting that the brain operates in a nonequilibrium
stationary state. Furthermore, the level of nonequilibrium, measured by entropy production or irreversibility appears to be a crucial signature of cognitive
complexity and consciousness. The subsequent study of neural dynamics
from the perspective of nonequilibrium statistical physics is an emergent field
that challenges the assumptions of symmetry and maximum-entropy that
are common in traditional models. In this review, we discuss the plethora
of exciting results emerging at the interface of nonequilibrium dynamics and
neuroscience. We begin with an introduction to the mathematical paradigms
necessary to understand nonequilibrium dynamics in both continuous and
discrete state-spaces. Next, we review both model-free and model-based
approaches to analysing nonequilibrium dynamics in both continuous-state
recordings and neural spike-trains, as well as the results of such analyses.
We briefly consider the topic of nonequilibrium computation in neural systems,
before concluding with a discussion and outlook on the field.
\end{abstract}


\begin{document}
		
    \maketitle 
    \thispagestyle{firststyle} \tauabstract 
    \tableofcontents
    
\nolinenumbers
\section{Introduction}
\label{sec: intro}
\begin{figure*}
    \centering
    \includegraphics[width=\linewidth]{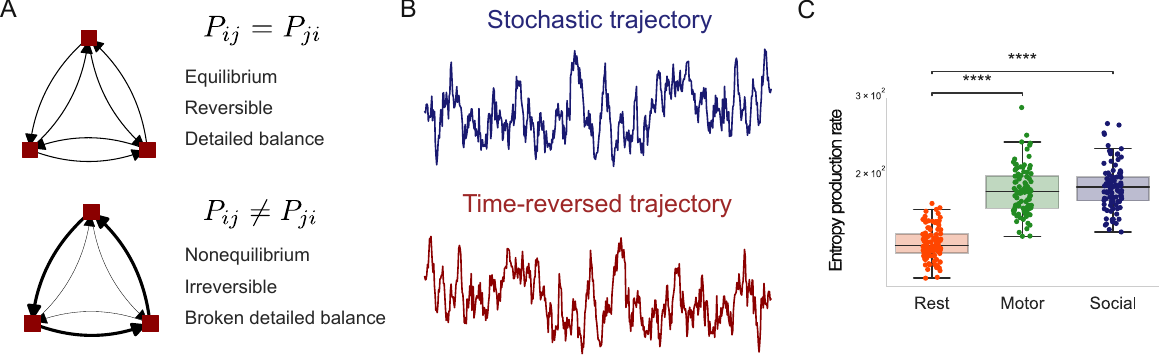}
    \caption{\textbf{Nonequilibrium brain dynamics.} A. Nonequilibrium dynamics are characterised by stationary probability currents, which occur when joint transition probabilities contain asymmetry i.e. $P_{ij}\neq P_{ji}$, that cause a system to violate detailed balance and produce irreversible trajectories. B. To estimate irreversibility, one must compare the statistical properties of a stochastic trajectory with its time-reversal
. C. When this is applied to neural recordings, differences between experimental conditions appear, such as an increase in irreversibility and entropy production in task compared to rest. Adapted from \cite{nartallokaluarachchi2024broken}.}
    \label{fig: nonequilibrium}
\end{figure*}

\taustart{I}nformation processing in human and animal brains is based on complex and coordinated neural activity that facilitates computation, cognition, and consciousness. The ability to record neural activity over time, through an array of neuroimaging and electrophysiological modalities, allows for the analysis and modelling of such spatiotemporal dynamics in an attempt to unravel the relationship between neural activity and cognition. Within this paradigm, dynamical systems, particularly those formulated as networks of discrete, interacting variables, have become the primary abstraction for the modelling of neural dynamics across an array of time and length-scales \cite{gerstner2014neuronaldynamics,Izhikevich2006dynamical,basset2017networkneuro}. Models range from the spiking behaviour of individual neurons, such as the seminal Hodgkin-Huxley model \cite{Hodgkin1952membrane}, to the mean-field dynamics of neural masses and oscillators \cite{Coombes2023neurodynamics}. Such abstractions have lent the language and results of dynamical systems, network theory, statistical physics and complex systems to the computational neuroscientist, prompting the analysis of emergent phenomena in the dynamics of brain networks \cite{chialvo2010emergent,Sporns2022complex,papo2024braincomplexnetwork}.\\\\
A particular area of `neurophysics' that has only recently garnered significant attention is the \textit{nonequilibrium} nature of neural dynamics. Discussion of nonequilibrium dynamics begins with the famed \textit{second law of thermodynamics}, which dictates that the average entropy of an isolated system increases as time flows forwards \cite{Fermi1937thermo}. More colloquially, it asserts that, over time, systems tend to disorder and, ultimately, thermodynamic equilibrium, so-called `heat death', where no energy (or information) is dissipated between agent and environment. Beginning with Schrödinger, it has long been argued that a fundamental characteristic of living systems is their ability to resist thermodynamic equilibrium by maintaining themselves in a nonequilibrium state \cite{Schrodinger1944whatislife,fang2019nonequilibrium,Gnesotto2018brokendetailedbalance}. In contrast with equilibrium dynamics, nonequilibrium dynamics are time-irreversible and thus induce the so-called `arrow of time' \cite{eddington1928}. Furthermore, in physical systems, they are characterised by an input of energy, the dissipation of heat to their environment, and the associated production of entropy. Key theoretical and experimental advancements in statistical mechanics have led to an expansive mathematical framework for the treatment of small nonequilibrium systems with appreciable fluctuations, described by so-called \textit{stochastic thermodynamics} \cite{seifert2012thermodynamics,Ciliberto2017stochastic}. These techniques have further lent themselves to the analysis of biological data, confirming the ubiquity of nonequilibrium dynamics in micro- and mesoscopic biological systems \cite{Gnesotto2018brokendetailedbalance,brangwynne2008cytoplasmicdiffusion,yin1999nonequilibriumRNA,Huang2003ecoli,mehta2012cellularcomputation,stuhrman2012cytoskeleton,battle2016brokendetailedbalance,diterlizzi2024variancesum}.\\\\
Whilst thermodynamic quantities such as energy and heat are well defined in microscopic biological dynamics, the same cannot be said of the macroscopic level, such as that of physiological recordings. Nevertheless, such recordings can still be analysed through the lens of nonequilibrium dynamics. In particular, one can directly measure the \textit{irreversibility} of the dynamics, which is a statistical quantification of the degree to which \textit{`forward'} trajectories diverge from their time-reversal. This analysis shows that nonequilibrium, irreversible dynamics are also ever-present in macroscopic physiological processes such as human heartbeats \cite{peng2019nonequilibrium,costa2005heart}, large-scale neural dynamics \cite{lynn2021detailedbalance} and even collective dynamics such as evolution \cite{England2013selfreplication} and social learning \cite{vaidya2021sociallearning}. However, without the physical interpretation of thermodynamics, irreversibility no longer corresponds to the dissipation of heat or the consumption of energy which renders the presence of nonequilibrium dynamics alone, uninformative. Nonetheless, the level of nonequilibrium has been shown to vary significantly between system states, such as increased reversibility in arrhythmic heartbeats \cite{costa2005heart} and impaired brain dynamics \cite{guzman2023impairedconscious,cruzat2023alzheimers}. Fig. \ref{fig: nonequilibrium} highlights a typical approach to analysing such systems. Nonequilibrium processes violate the \textit{detailed balance condition}, which is met when probability fluxes vanish the steady-state, and thus produce time-irreversible dynamics. As a result, these stochastic trajectories, captured by biological recordings, can be compared with their time-reversals to measure differences and quantify irreversibility. This can be used to find variation between experimental conditions, but is also a way to characterise the statistical properties of a time series. Crucially, this suggests that irreversibility is a key emergent feature of an underlying nonequilibrium state where equilibrium techniques could conceal important features of the system under consideration. Furthermore, despite being macroscopic in nature, neural dynamics remain the output of a biologically-embedded system which is subject to laws of thermodynamics. Understanding the multi-scale cascade that begins with cellular thermodynamics, and becomes `coarse-grained' at the level of macroscopic dynamics remains an interesting but challenging problem.
\\
\\
In this review, we focus on nonequilibrium steady-states in the macroscopic dynamics of the brain, presenting both a technical and interpretative commentary on a developing area of research. Our intended audience is both the mathematician/physicist interested in applying techniques from stochastic nonequilibrium dynamics to neuroscientific data, as well as the neuroscientist looking for a comprehensive, yet accessible, introduction to the mathematical framework and data analytic techniques that are available. Although the term \textit{`thermodynamics'} has previously been used in reference to both theoretical and biological computation \cite{kringelbach2024thermodynamics,Wolpert2019thermodynamics}, in the absence of a physical notion of `heat' or `energy' in neural dynamics, we will refrain from using the term, but instead refer to \textit{nonequilibrium dynamics} with a few exceptions in Sec. \ref{sec: nonequilibrium neural computation}.\\\\
This review is organised as follows. First, we motivate the study by recapping recent results which highlight nonequilibrium measurements as a signature that varies across both conscious state and cognitive task. Second, we highlight prominent modelling and analysis techniques that hinge on equilibrium assumptions. Third, we introduce the mathematical preliminaries required for analysing nonequilibrium dynamics, beginning with continuous-space processes followed by discrete-state processes. Next, we consider both model-based and model-free techniques for the analysis of continuous, then discrete, neural data. Penultimately, we offer a brief discussion of adjacent areas of neuroscience that investigate nonequilibrium dynamics within the context of computation. Finally, we present a discussion and outlook on the field.
\subsection{Neurons, populations and neural dynamics}
\label{sec: neurons}

\begin{figure*}
    \centering
    \includegraphics[width=\linewidth]{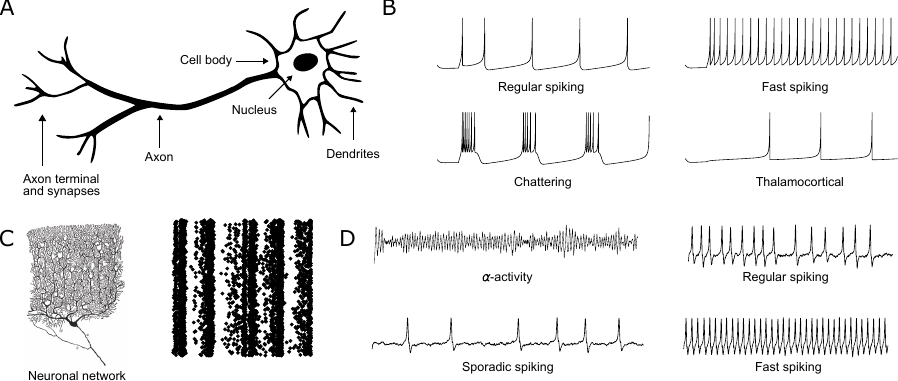}
    \caption{\textbf{Neurons, populations and dynamics}. A. A schematic diagram of a single neuron. Adapted from Notjim by CC BY-SA 3.0 (\url{https://commons.wikimedia.org/w/index.php?curid=4824168}). B. Action potentials from different dynamic regimes. These are simulated using the Izhikevich model \cite{Izhikevich2006dynamical}. C. Neurons are organised into networks of connected units, which produce collective activity. This can be visualised as a spike-train in a raster plot. Drawing by Santiago Ramón y Cajal (Public Domain). D. The \textit{average} activity of a population or network or neurons can be approximated with a mean-field model. These models have an array of dynamic regimes. These are simulated from the Jansen-Rit neural-mass model \cite{Coombes2023neurodynamics}.
    }
    \label{fig: neurons}
\end{figure*}

It has long been known that the brain is the seat of cognition for both humans and many other animals, as such it was of profound interest to early philosophers and scientists alike. It was not until the work of Golgi and y Cajal, that \textit{neurons} were identified as the principals computational unit of  the brain \cite{Golgi1873sulla,Cajal1888estructura}. Despite the zoo of different cell types in the brain, the study of cognition has since focused primarily on neurons and their connectivity.\\
\\
Briefly, neurons communicate via electrical signals that propagate along their \textit{axons} (Panel A, Fig. \ref{fig: neurons}). They operate as threshold units, requiring a sufficient input of current to prompt a burst of activity known as an \textit{action potential} or \textit{spike} that travels along the axon. Hodgkin and Huxley were able to derive the equations for neuronal spiking from the dynamics of ionic channels in the neuron \cite{Hodgkin1952membrane}, thus allowing for the modelling of neuronal activity. These models take the form of ordinary differential equations (ODEs) and have a range of dynamic regimes with different spiking behaviour (Panel B, Fig. \ref{fig: neurons}). Through \textit{synapses}, neurons are connected and action potentials from incoming connections serve as input current for a post-synaptic neuron, thus potentially inducing further spikes. In this fashion, populations of neurons exhibit coordinated, collective behaviour. The activity of a neuronal network can be simplified by discretizing the spikes as discrete threshold events, ignoring the continuous dynamics of the action potential (Panel C, Fig. \ref{fig: neurons}). The dominant \textit{connectionist} paradigm in neuroscience argues that cognition in the brain is performed in a distributed fashion amongst networks of simple processing units, and has inspired the development of artificial neural networks and deep learning \cite{lecun2015deeplearning}.\\
\\
Recording from individual neurons in the brain is challenging and requires invasive electrophysiology. As such, these modalities primarily focus on small animals such as mice and rats, with limited data for humans. On the other hand, neuroimaging and encephalography methods, such as functional magnetic resonance imaging (fMRI), magnetoencephalography (MEG) and electroencephalography (EEG), are able to non-invasively capture large-scale activity in the brain. These modalities record proxies of neural activity such as blood-flow into brain regions (fMRI), magnetic fields measurable from outside the head (MEG), and electrical potentials on the scalp (EEG). As a result, these data can be considered to be the \textit{average} activity in a region of interest (ROI) of the brain, known as a \textit{neural-mass}. This neural-mass is composed of a large number of interacting neurons with heterogenous connectivity and cell-types. Nevertheless, for the purposes of modelling such data, we typically opt for a \textit{mean-field} approach, where a single model is used to describe the average activity of a entire neural population. A diverse range of \textit{neural-mass models} exist, with a variety of complexity, and they are able to reproduce an array of different activity types, in alignment with neuroimaging data \cite{Coombes2023neurodynamics} (Panel D, Fig. \ref{fig: neurons}). By considering macroscopic connections between neural masses, white-matter fibres visible under diffusion tensor imaging (DTI) \cite{Hagmann2008mapping}, we can model the collective behaviour of the whole brain, constrained by this connectivity, up to the resolution of the data at hand.\\
\\
Both the connectionist perspective on neuronal activity, and the resolution of neuroimaging, encephalography and electrophysiological data, dictate the level of abstraction from which we attempt to understand neural systems. Modelling the brain as an interconnected network of discrete interacting units is the dominant paradigm that is most aligned with experimental data - although this is performed at a diversity of scales, including both discrete and continuous outputs. This naturally sets the scene for a mathematical paradigm that focuses on nonlinear dynamics, network science, statistical physics, and time-series analysis. 
\subsection{Nonequilibrium signatures of consciousness and cognition}
\label{sec: signatures}
The dynamics of large-scale brain activity, as recorded with functional neuroimaging, exhibit broken detailed balance and temporal irreversibility \cite{lynn2021detailedbalance}. However, it is not immediately obvious why such an emergent phenomenon should occur or, separately, why it is relevant to the dynamics of neural systems. Naturally, the trajectory of a single action-potential is time-irreversible which suggests that the macroscopic dynamics should exhibit time-irreversibility. However, the effect of coarse-graining, from the micro- to the macro-scale of a nonequilibrium system, is complex. Notably, coarse-graining can disguise an underlying nonequilibrium system as an equilibrium one \cite{Esposito2012coarsegraining,Egolf2000equilibrium}.\\\\
The crucial insight motivating the study of nonequilibrium dynamics in the brain is the variation of the \textit{degree of nonequilibrium}, as measured by irreversibility or rate of entropy production, across both conscious state, cognitive task or disease condition. A range of studies, each employing a distinct measure of nonequilibrium, have found that the irreversibility of large-scale neural activity during cognitive tasks is higher than in a resting state \cite{lynn2021detailedbalance,deco2022insideout,deco2023tenet,deco2023violations,nartallokaluarachchi2024broken,bolton2023AoT}. It is argued that the level of nonequilibrium increases with the demands of the cognitive task, with the resting state being the most reversible. However, due to the diversity of methods used, there is no consensus ranking of the seven tasks from the Human Connectome Project (HCP) database \cite{vanessen2013hcp} used in each of these studies. This highlights the difficulty of measuring nonequilibrium in high-dimensional data and the dependence of results on the nuances and assumptions of each individual method.\\
\\
In addition to task-based activity, the level of nonequilibrium varies across levels of consciousness with decreased entropy production and irreversibility during sleep \cite{gilson2023OU,sanzperl2021nonequilibrium}, under anaesthetic drugs \cite{deco2022insideout,delafuente2023irreversibility}, under psychedelics \cite{Shinozuka2024LSD,Deco2024psilodep} and in diseases of consciousness such as minimally conscious state and unresponsive wakefulness state \cite{guzman2023impairedconscious}. Finally, it has been shown to vary across disease states with altered levels of irreversibility in Alzheimer's disease \cite{cruzat2023alzheimers}, bipolar disorder \cite{bernardi2023ocd,zanin2020pathology}, Schizophrenia \cite{deco2023tenet,zanin2020pathology}, attention deficit hyperactivity disorder \cite{deco2023tenet}, Parkinson's disease \cite{zanin2020pathology} and epilepsy \cite{Schindler2016epilepsy,donges2012visibility,zanin2020pathology,martinez2018ordinal}.\\\\
Beyond comparisons on the overall degree of irreversibility, nonequilibrium techniques have shown promise as a more general modelling and analysis tool. For example, time-lagged correlations, inspired by analysis of irreversibility, outperform functional connectivity in the classification of brain states \cite{tewarie2023nonerev}. Additionally, irreversibility information has been used to improve whole-brain models of large-scale brain activity by tuning effective connections \cite{deco2023violations,kringelbach2023movie,kringelbach2024thermodynamics,Deco2024psilodep}. Moreover, irreversibility has been used to identify causal flows \cite{bolton2023AoT} and higher-order interactions \cite{nartallokalu2025multilevel,lynn2022decomposing,lynn2022localAoT} in neural activity. Finally, nonequilibrium dynamics have been linked to the asymmetric hierarchical organisation of the human brain and used to infer hierarchies of brain regions from neuroimaging time-series \cite{nartallokaluarachchi2024broken,Deco2024psilodep,kringelbach2024thermodynamics}.

\subsection{Comments on consciousness and its correlates}

Before moving on from the previous section, it is worth clarifying our use of the often-controversial term \textit{consciousness}. In simple terms, being conscious refers to the subjective phenomenon of `having an experience'. Despite broad agreement that we are referring to the same concept, consciousness remains an illusory and vague concept that so far has eluded a satisfactory rigorous scientific definition. Nevertheless, developing a mathematical or scientific theory of consciousness is often described as one of the most important, interesting and challenging problems of science and philosophy, especially by the people who write extensively about it.\\
\\
Despite many valiant efforts,  the study of consciousness is, as of yet, far from being an empirical science due to the challenges of developing  consensus definitions or relating theory to empirical analysis. Moreover, the paradox of the `hard problem of consciousness', as posited by Chalmers \cite{Chalmers1995hardproblem}, suggests that even a complete neuroscientific understanding of consciousness may fail to give a satisfactory answer to the more esoteric question of why we have experiences at all.\\
\\
Despite a clear media appeal, such discussions are beyond the scope of our current mathematical and physical frameworks, and we make no claims about the relationship between nonequilibrium physics and the hard problem. Nevertheless, neuroimaging collected from participants in a diverse array of \textit{brain states} can give way to markers that indicate the level of \textit{activity} or \textit{awareness} that are usually conflated with the notion of consciousness. For example, most would agree that a cadaver, an anaesthetised person, a sleeping person, and an awake person can be placed on a scale of increasing levels of consciousness, even if this appears impossible to objectively confirm. Proceeding under this assumption, we can probe the neural dynamics underlying these states to investigate dynamical features that display a correlation with the assumed ladder of conscious states. These so-called `neural correlates of consciousness' \cite{Koch2016neural}, which we refer to in the proceeding section, make no direct headway towards the hard problem of consciousness, yet may give us insight into the dynamical features that appear to orchestrate conscious behaviour.\\
\\
It is from this vantage point that we make reference to the EPR or level of nonequilbrium as a signature of conscious behaviour \cite{delafuente2023irreversibility}, arguing that any more direct claims about the application of nonequilibrium physics to an elusive `theory of consciousness' goes beyond the evidence collected hereto. 
\subsection{Equilibrium models of neural activity}
\label{sec: Equilibrium models of neural activity}
Despite the many results highlighted in the previous section, equilibrium models of neural dynamics remain the dominant framework. Whilst some such models explicitly make the assumption of an equilibrium stationary state, many models implicitly assume time-reversible dynamics and reciprocal interactions. In this section, we highlight where these equilibrium assumptions are present in widely used techniques.
\subsubsection{The equilibrium Ising model}
\label{sec: intro ising}
In a pair of seminal papers, E.T. Jaynes introduced the \textit{principle of maximum entropy} \cite{Jaynes1957I,Jaynes1957II}. In simple terms, the principle argues that given empirical data and a set of constraints to which the model/distribution which generated the data much adhere to, the distribution which satisfies the constraints yet maximises the entropy/uncertainty is the preferred choice \cite{presse2013maxent}. A classic example is to consider the binary activity of a set of neurons, $(s_1,...,s_N)(t_k)$, each being active $(+1)$ or inactive $(-1)$ at each time-step $t_k$. From observed data in the form of neural spike-trains, one can calculate both the average firing probability of each neuron, $\langle s_i\rangle$, as well as the pairwise correlations,  $\langle s_is_j\rangle - \langle s_i\rangle\langle s_j\rangle$. Under these constraints, the distribution that maximises the entropy is precisely the \textit{Boltzmann distribution},
\begin{align}
    P(s_1,...,s_N)=\frac{1}{Z}\exp\left[ \sum_i h_is_i + \frac{1}{2}\sum_{i\neq j}J_{ij}s_is_j\right], \label{Eq: Boltzmann}
\end{align}
where $Z$ is a normalising factor known as the \textit{partition function} and $\{h_i,J_{ij}\}$ correspond to Lagrange multipliers that one seeks to infer in order to match the statistics $\langle s_i\rangle, \langle s_is_j\rangle$ between model and data \cite{schneidman2006weak,bialek1996spikes}. Moreover, this model assumes that $J_{ij}=J_{ji}$. It is important to note that the entropy of this distribution is not equivalent the entropy production rate associated with nonequilibrium dynamics. This distribution corresponds identically to the famous \textit{Ising model} of magnetic spins where $\{h_i\}$ are the external fields to each spin and $\{J_{ij}\}$ are the symmetric pairwise coupling strengths \cite{Nishimori2001statphys,gasper2018ising}. The Boltzmann distribution  (\ref{Eq: Boltzmann}) corresponds to the \textit{equilibrium steady state} of an Ising model with symmetric pairwise couplings and satisfies the fluctuation-dissipation theorem, has zero entropy production and is time-reversible, rendering it unable to model nonequilibrium activity in neural spiking \cite{deco2023violations}. This equilibrium Ising approach has been used extensively to study the structure of neural spike-trains \cite{meshulam2017collective,meshulam2023successes,lynn2024solvable,lynn2024minimax,tang2008maxent,shlens2006primate,schneidman2006weak,Rosch2024spontaneous} with a particular focus on emergent phenomena from statistical physics such as criticality \cite{tkacik2015thermodynamics} and `crackling-noise' \cite{poncealvarez2018crackling}. Similarly, Hopfield and Boltzmann networks use symmetric couplings to construct neural networks with equilibrium dynamics that are defined by the gradient of a scalar potential, where they are able to encode memories as energy minima \cite{hopfield1982hopfield,ackley1985boltzmann}. In lieu of a nonequilibrium counterpart, the use of the equilibrium Ising model is a conscious choice justified in order to leverage the existing techniques for studying emergent phenomena in the Ising model. Whilst useful, this model of neural computation neglects the nonequilibrium dynamics that emerge from biologically-realistic asymmetric couplings \cite{yan2013landscape}.
\subsubsection{Symmetric whole-brain models}
\label{sec: symmetric whole brain}
\begin{figure*}
    \centering
    \includegraphics[width=\linewidth]{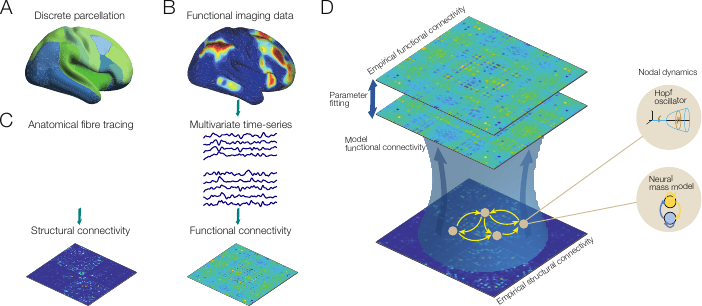}
    \caption{\textbf{Symmetric whole-brain modelling.} A. In order to model the brain, we select a parcellation into discrete brain areas. B. From functional imaging data, such as fMRI, nodal activity in the form of multivariate time-series can be extracted. From these data, we obtain a functional connectivity matrix. C. Anatomical fibre tracing, such as diffusion tensor imaging, allows the reconstruction of connectivity between brain areas. This yields the structural connectivity matrix. D. A whole-brain model is then defined by selecting a network dynamical system, such as a neural mass model or Hopf oscillator network. This model is constrained by the empirical structural network. The parameters of each regional dynamic model are fit by minimising the difference between the functional connectivity simulated from the model, and that calculated from the empirical data. Adapted from \cite{patow2024wholebrain}.}
    \label{fig: symm whole brain}
\end{figure*}
In the modelling and analysis of large-scale continuous recordings, similar equilibrium assumptions are pervasive and stem from both oversight and restrictions implicit in the available data.\\
\\
The traditional paradigm for modelling of the whole-brain activity is illustrated in Fig. \ref{fig: symm whole brain}. We begin with a \textit{parcellation} of the brain into discrete regions. This allows for functional imaging data to be converted into a real-valued \textit{multivariate time-series} (MVTS), $\{(\hat{x}_1,...,\hat{x}_N)(t_k)\}_{k=1}^T$, which one attempts to model using a network dynamical system,
\begin{align}
\Dot{x_i}&=f_i(x_i)+\sum_{j}A_{ij}g_{ij}(x_i,x_j),
\end{align}
where $x_i(t)$ represents the activity of region $i$ over time, $f_i(\cdot)$ represents a nodal dynamic, $\bm{A}=(A_{ij})$ represents the pairwise regional couplings and $g_{ij}(\cdot)$ represents a coupling function \cite{patow2024wholebrain,deco2014greatexpectations}. Typically, regions are assumed to be homogeneous ($f=f_i, g = g_{ij}$), with regional dynamics given by neural-mass or oscillatory models \cite{Coombes2023neurodynamics}, such as the \textit{Hopf} \cite{deco2017resting} or \textit{dynamic mean field} models \cite{Kringelbach2020DMF}. The interaction network $\bm{A}$ is the \textit{structural connectivity matrix} (SC) between brain regions, usually a participant-averaged network obtained separately from the same functional imaging that generated the time-series \cite{deco2014greatexpectations}. Although in animals invasive techniques such as tract-tracing can measure directed, asymmetric connections between brain regions \cite{kotter2004online}, no such data is directly available for human brain networks, so-called \textit{connectomes}, which rely on non-invasive imaging techniques \cite{Hagmann2008mapping}, which lead to undirected representations of brain networks ($A_{ij}=A_{ji}$, $\bm{A}=\bm{A}^{\top}$) despite the fact that they are inherently directed \cite{kale2018directed}.\\
\\
The dynamics of directed networks differ from their undirected counterparts in many ways \cite{asllani2018nonnormal,johnson2020digraphs,krakauer2023brokensymmetry}, but most relevantly, they typically produce time-irreversible, nonequilibrium dynamics \cite{nartallokaluarachchi2024broken}. In order to model the nonequilibrium dynamics present in human neuroimaging, the asymmetries in interaction strengths must be inferred directly from the MVTS in order to obtain a so-called \textit{effective network} \cite{friston2011FCEC}. However, inferring a whole-brain model whilst assuming knowledge of the SC is already a difficult challenge due to large parameter space and the relatively short length of the data. Attempting to do this whilst simultaneously inferring the network becomes an almost impossible, and potential under-determined \cite{prasse2022predicting}, global optimisation challenge that must be circumvented with assumptions and heuristics.\\
\\
In addition to structural symmetry, whole-brain models often aim to match the \textit{functional connectivity matrix} (FC) of the model and the data, as shown in Panel D of Fig. \ref{fig: symm whole brain}. The FC matrix, $\bm{C}=(C_{ij})$,is typically given by,
\begin{align}
    C_{ij} = r(x_i,x_j),
\end{align}
where $r(x,y)$ represents the \textit{Pearson correlation coefficient} (PCC) between the series $x$ and $y$ \cite{friston2011FCEC}. As a result, the FC is typically symmetric ($C_{ij}=C_{ji}$, $\bm{C}=\bm{C}^{\top}$) and therefore incapable of capturing asymmetries in interaction strengths. Between the symmetric models and the symmetric summary statistics, connectome-based whole-brain models typically fail to capture the nonequilibrium dynamics of the underlying data. As a result, one must turn to simplified or statistical models such as those described in Sec. \ref{sec: continuous-valued recordings}.\footnote{It is worth noting that whilst PCC is the standard metric, FC can be defined using any pairwise metric of which there are many, including a range of directed metrics \cite{cliff2023pairwise}. It is unclear if symmetric models can fit asymmetric summary statistics.}
\subsubsection{Directedness in non-human connectomes}
\label{sec: directed}
Whilst SC in the human brain is typically restricted to undirected representations, invasive methods have reconstructed directed connectomes for animal and insect brains \cite{kotter2004online,kale2018directed}. For example, the \textit{FlyWire} project recently published a complete, directed wiring-diagram of the \textit{Drosophila} brain \cite{Dorkenwald2024flywire}. Its directed structure reveals that some neurons operate as \textit{`broadcasters'}, with many more out-going connections than incoming, whilst others are \textit{`integrators'} with a larger number of incoming connections \cite{Lin2024networkstatsfly}. Moreover, the density of reciprocal and non-reciprocal connections vary significantly throughout the fly brain. Simulated dynamics on the asymmetric fly brain shows violation of the fluctuation-dissipation theorem, and the emergence of nonequilibrium dynamics \cite{Odor2025fluctuation}.\\
\\
In addition to the fly, directed connectomes are available for the macaque \cite{kotter2004online}, cat \cite{Scannell1999cat}, mouse \cite{Dong2008allenmouse} and \textit{C. Elegans} \cite{Varshney2011celegens}. Directionality has significant impact on the structural properties of these networks such as degree-distribution, clustering coefficient, small-world index and scale-free properties \cite{kale2018directed, Cirunay2025scalefree}, with \textit{rich-club} regions showing higher in-degree than out-degree \cite{kale2018directed,dereus2013cat}. Finally, directed structure can constrain dynamics with a strong correlation being shown between in-degree characteristics and resting-state fMRI in the mouse \cite{Sethi2017structural}.\\
\\
In the human, directed connectomes have been restricted to functional and effective networks computed using simple models or directed correlation measures \cite{Rubinov2010complex}. More recently, there has been renewed interest in identifying directed connections in the human brain from recordings of activity \cite{Greaves2025structurally}. These approaches include autoregressive models \cite{Seguin2019directionality}, `causal' models \cite{Friston2009causal} and deep learning \cite{Ji2023surveyeffective} amongst others.\\\\
More generally, structural analysis of connectomes has been considered from a range of network-theoretic perspectives including modularity \cite{Meunier2010modularity}, discrete curvature \cite{Farooq2019networkcurvature,Elumalai2022ricci}, fractal geometry \cite{Ansell2024fractal}, with links to dynamics being realised through network control theory \cite{Gu2015controllability,Kim2020linear,Parkes2024networkcontrol}.
\subsection{Relating model properties with empirical data}
\label{sec: properties}

Throughout this review, we consider abstract models, typically originating from the modelling of physical systems, and apply them to neuroscientific data. It is worth taking a moment to consider the relationship between the properties of a mathematical model and the data which it attempts to describe. To begin with, we note that within the paradigm of statistical physics, such a problem runs rampant. This problem is a natural consequence of the development of statistical physics to describe physical systems such as gases, magnets or fluids, only to find that the same principles apply to a range of systems, both abstract and (bio)physical. A classical example is the relationship between Gibbs' notion of entropy for a collection of classical particles \cite{Gibbs1902statmech} and Shannon's information-theoretic entropy \cite{Shannon1948theory}. The first is a thermodynamic, and therefore physical, property of the system which describes the number of microstates consistent with a collection of macroscopic thermodynamic properties. In Shannon's form, it can readily be applied to abstract systems to describe the uncertainty in a random variable given its probability distribution. Shannon's entropy is precisely the same as Gibbs' when the random variable describes the energy of the microstates - which need not be the case.\\
\\
In a similar vein, throughout statistical physics approaches to neuroscience or other complex systems, one may make reference to the \textit{temperature}, \textit{energy}, and \textit{entropy production rate} of a system, without necessarily referring to the physical properties of the system, but instead describing the probabilistic behaviour of its dynamics at a particular coarse-grained scale \cite{Advani2013statmech,Barkley2016econophysics}. This distinction is most important when describing a system at an abstract level, but which is physically instantiated in the world. For example, whilst we may describe the `temperature of activity in the brain' through an analogy with the Ising model, this may or may not be related to the physical temperature of the brain in real terms - the relationship is entirely unclear. Throughout this review, we focus on the EPR of neural systems, in reference to the irreversibility of neural dynamics and the presence of dissipative forces in phase-space. The relationship between the EPR of a neural model or data and its physical properties is not apparent. The thermodynamic properties of \textit{information processing} \cite{Wolpert2019thermodynamics, parrondo2015information} complicate this analogy even further, by introducing another notion of entropy production for computations - including those being performed in the brain. With the exception of Sec. \ref{sec: nonequilibrium neural computation}, we refer to the abstract EPR associated with the dissipative phase-space dynamics, and venture no claims about the energetic costs of either the brain as an organ, or the brain as a computer.\\
\\
A closely connected problem is determining when data can be said to take on the property of a dynamic model. For example, when a model that exhibits criticality, chaos, or turbulence - each a well-defined concept in mathematical and physical systems - is fit to neural data, where these phenomena are less well-defined, is it fair to say that the brain is then critical, chaotic, or turbulent? It remains a point of contention. When addressing a similar topic, Bassett et al \cite{Bassett2018nature} introduce three notions of validity for a network-based model of the brain, which we adapt here to notions of neural dynamics.
\begin{enumerate}
    \item \textit{Descriptive validity}: Does the model produce trajectories that resemble the statistical properties of the data?
    \item \textit{Explanatory validity}: Does the model provide evidence for a causal mechanism in the system? By extension, can it be used to test causal relationships on the basis of this model?
    \item \textit{Predictive validity}: Can the model be used to make predictions that are later confirmed by empirical data?
\end{enumerate}
It is clear from these definitions that the lowest barrier to entry for a neural model is descriptive validity which simply describes whether or not a model can be tuned to fit neural data with sufficient accuracy. In this case, attributing properties of the model to the data is the least relevant and supported. In our context, this could be related to the finding that neural dynamics exhibit significant\footnote{See Sec. \ref{sec: Testing for significance, stationarity and bias in nonequilibrium measures} for a discussion of significance testing for time-irreversibility.} irreversibility, and are therefore best described by nonequilibrium models. On the other hand, linking nonequilibrium dynamics to the hierarchical asymmetric structure of effective brain networks then provides a causal mechanism for the violation of detailed balance \cite{nartallokaluarachchi2024broken}. Similarly, the leveraging of nonequilibrium dynamics for sequencing in associative memory models \cite{yan2013landscape} is also a causal, explanatory mechanism. Finally, showing that the level of nonequilibrium varies between brain states allows for predictive validity, where nonequilibrium signatures can be used to classify unlabelled samples from neuroimaging data at a variety of consciousness levels.\\
\\
In the case of criticality, chaos, or turbulence, a similar analysis can be made. If the optimal working point of the model coincides with the specific regime associated with a property, this implies descriptive validity. The explanatory validity depends on the causal mechanism associated with the property i.e. optimal sensitivity to information for critical dynamics. Ideally this hypothesis is made apriori, rather than an aposteriori explanation of observed results. Finally, if the causal mechanism allows for predictions to be made about the dynamics in unseen data, or the behaviour of the system in response to perturbations, it satisfies predictive validity.

\section{Mathematical preliminaries for nonequilibrium processes}
\label{sec: math prelim}
In this section, we introduce the mathematical preliminaries necessary for constructing nonequilibrium models and measures. We focus on stochastic dynamics in the steady-state for both continuous and discrete spaces. We begin with continuous-space processes in the form of Langevin equations which are relevant for the modelling and analysis of continuous neural recordings such as neuroimaging and electro-physiological signals. Next we consider discrete-state processes which are relevant for the modelling of neural spike-trains or analysing state-space models of neural data.
\subsection{Continuous-space processes and Langevin equations}
\label{sec: continuous space}
We focus on continuous stochastic dynamics expressed by the over-damped Langevin equation \cite{Kampen1992},
\begin{align}\label{eq: langevin}
    \dot{\mathbf{x}}(t) & = \bm{F}(\mathbf{x},t) + \bm{\xi}(\mathbf{x},t),
\end{align}
where $\mathbf{x}(t)\in \mathbb{R}^n$ is the state of the system at time $t$, $\bm{F}$ models the deterministic, driving force of the system, which known as the \textit{drift}, and $\bm{\xi} \in \mathbb{R}^n$ are Gaussian fluctuations, known as the \textit{diffusion} or \textit{noise}. The diffusion is specified by its auto-correlation function which is expressed as,
\begin{align}
    \langle \bm{\xi}(\mathbf{x},t),\bm{\xi}(\mathbf{x},t')\rangle = 2\bm{D}(\mathbf{x})\delta(t-t'),
\end{align}
implying that diffusion is uncorrelated in time. Whilst this is the most general form, common assumptions include that the drift is constant in time, $\bm{F}(\mathbf{x},t)\equiv \bm{F}(\mathbf{x})$, or that the diffusion is spatially-homogeneous, $\bm{\xi}(\mathbf{x},t) \equiv \bm{\xi}(t)$, and thus can be represented by the constant matrix $\bm{D}$, or that $\bm{D}(\mathbf{x})$ is a diagonal matrix. In the following, we will assume that $\bm{F}(\mathbf{x},t)\equiv \bm{F}(\mathbf{x})$. Whilst individual trajectories are stochastic and unpredictable, the \textit{probability density} of trajectories evolve according to the \textit{Fokker-Planck} (FP) equation\footnote{This known as a \textit{partial differential equation} (PDE) as it describes the relationship between the different derivatives of a multivariate function.},
\begin{align}\label{eq: FP}
    \frac{\partial P(\mathbf{x},t)}{\partial t} = &-\sum_i \frac{\partial}{\partial x_i} \left [ F_iP\right] + \sum_{i,j} \frac{\partial}{\partial x_i \partial x_j} \left[ D_{ij}P\right],
\end{align}
where $P(\mathbf{x},t)$ is the probability of a trajectory having the value $\mathbf{x}$ at time $t$ \cite{Kampen1992}. This can be equivalently expressed as a conservation law,
\begin{align}
\frac{\partial P}{\partial t} + \nabla\cdot \bm{J} & = 0,
\end{align}
where $\bm{J}(\mathbf{x},t)$ is the \textit{probability flux} defined as,
\begin{align}
    \bm{J} & = \bm{F}P - \nabla\cdot (\bm{D}P),\\
    &= (\bm{F}-\nabla \cdot \bm{D})P - \bm{D}\nabla P,
\end{align}
where $\nabla\cdot$ represents the divergence in space and $\nabla$ is the gradient. Over time, the system can converge to a \textit{steady-state} specified by a stationary density, $P_{\text{ss}}(\mathbf{x})$, which no longer depends on time. The stationary probability flux therefore has zero divergence, $\nabla \cdot \bm{J}_{ss} \equiv 0$. In an \textit{equilibrium steady-state} (ESS) the probability flux vanishes, $\bm{J}_{ss} \equiv 0$, which is the \textit{detailed balance} condition for SDEs. On the other hand, if $\bm{J}_{\text{ss}} \neq 0$, the steady-state contains probability currents, violates detailed balance, and is called a \textit{nonequilibrium steady-state} (NESS).\\
\\
A NESS is associated with a positive rate of entropy production. To define the entropy production rate we begin with the \textit{Shannon-Gibbs entropy} of a trajectory $\mathbf{x}(t)$ \cite{Jiang2004mathematical},
\begin{align}
    S = -\int_{\mathbb{R}^n}P(\mathbf{x},t)\log P(\mathbf{x},t)\; d\mathbf{x}.
\end{align}
Next, we differentiate with respect to time and use the FP equation to obtain,
\begin{align}
    \dot{S}(t) & = \int_{\mathbb{R}^n}(\log P(\mathbf{x},t)+1)\nabla \cdot \bm{J}(\mathbf{x},t) \; d\mathbf{x},
\end{align}
where we can then integrate by parts and apply that the flux vanishes at infinity (or the boundary),
\begin{align}
    \dot{S}(t) & = -\int_{\mathbb{R}^n}\nabla (\log P(\mathbf{x},t)+1)\cdot \bm{J}(\mathbf{x},t) \; d\mathbf{x},\\
    &= -\int_{\mathbb{R}^n}\frac{\nabla P(\mathbf{x},t)}{P(\mathbf{x},t)}\cdot \bm{J}(\mathbf{x},t) \; d\mathbf{x}.
\end{align}
From the expression for the flux, we can substitute $\nabla P / P$, which then yields,
\begin{align}
    \dot{S}(t) = &-\int_{\mathbb{R}^n}\bm{D}^{-1}(\mathbf{x},t)(\bm{F}(\mathbf{x},t)-\nabla \cdot \bm{D}(\mathbf{x},t))\cdot\frac{\bm{J}(\mathbf{x},t)}{P(\mathbf{x},t)}\\ &+ \int_{\mathbb{R}^n}\frac{\bm{J}^{\top}(\mathbf{x},t)\bm{D}^{-1}(\mathbf{x},t)\bm{J}(\mathbf{x},t)}{P(\mathbf{x},t)},\notag\\
    &= - h_d(t) + \Phi(t),
\end{align}
where $h_d$ is defined to be the average \textit{heat dissipation rate}, and $\Phi$ is the average rate of entropy production \cite{Jiang2004mathematical}. In the steady-state, $\dot{S}(t)=0$ which implies that $\Phi = h_d$ i.e. all entropy that is produced is dissipated to the environment as heat. Therefore in a NESS, the \textit{entropy production rate} (EPR), $\Phi$, is defined to be,
\begin{align}
    \Phi&= \int  \frac{\bm{J}_{\text{ss}}^{\top}\bm{D}^{-1}\bm{J}_{\text{ss}}}{P_{\text{ss}}}\;\;d\mathbf{x} \geq 0.
\end{align}
\subsubsection{The Helmholtz-Hodge decomposition}
\label{sec: HHD langevin}
\begin{figure*}
    \centering
    \includegraphics[width=\linewidth]{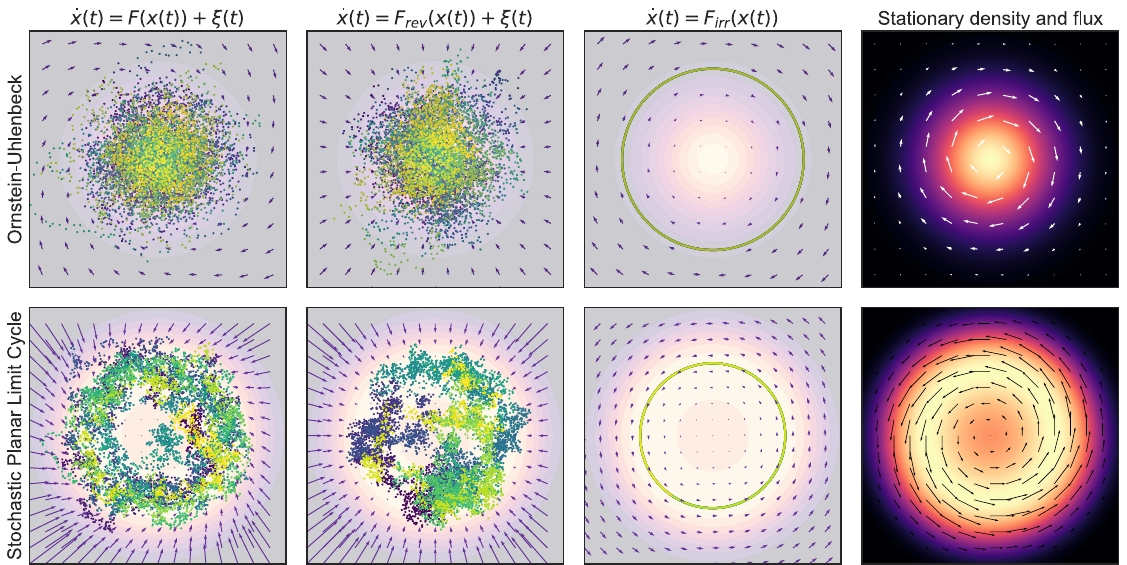}
    \caption{\textbf{The Helmholtz-Hodge decomposition.} Stationary processes admit a decomposition of their drift field, and by extension their trajectories, into irreversible and reversible components. The irreversible component drives rotation around the stationary distribution whilst the reversible component maintains the process at stationarity. The top row shows the decomposition of the Ornstein-Uhlenbeck process which is a linear process with a Gaussian stationary distribution. The bottom row shows the decomposition of the stochastic planar limit cycle which is a nonlinear process with a `Mexican-hat-type' stationary distribution. We show examples of decomposed trajectories depicted over the drift components (purple arrows) and the stationary density. On the right, we show the stationary density alongside the stationary probability flux (arrows). Details on the processes and decomposition are in Ref. \cite{nartallokaluarachchi2024decomposing}.}
    \label{fig: HHD}
\end{figure*}
Assuming the process is stationary, the drift can be expressed as,
\begin{align}
    \bm{F} &=  \frac{\bm{J}_{ss}}{P_{ss}} + \bm{D}\cdot \nabla \log P_{ss} + \nabla\cdot \bm{D},
\end{align}
which can be decomposed as follows,
\begin{align}
    \bm{F}& = \bm{F}_{\text{irr}} + \bm{F}_{\text{rev}},\\
    \bm{F}_{\text{rev}} &= \bm{D}\cdot\nabla \log P_{ss} + \nabla\cdot \bm{D},\label{eq: helmholtzsde}\\
    \bm{F}_{\text{irr}} &=\bm{J}_{ss}/P_{ss}, \label{eq: helmholtzsdeirrev}
\end{align}
where $\bm{F}_{\text{irr}}$ and $\bm{F}_{\text{rev}}$ represent the time-irreversible and time-reversible components of the drift respectively. In addition, $\bm{F}_{\text{irr}}$ is odd whilst $\bm{F}_{\text{rev}}$ is even under time-reversal \cite{DaCosta_2023}. As the drift can be expressed as the sum of a divergence-free flux, $\bm{J}_{ss}/P_{ss}$, and the diffusion-weighted gradient of a scalar potential, $U=-\log P_{ss}$, it is referred to as the \textit{Helmholtz-Hodge decomposition} (HHD) \cite{DaCosta_2023} because of its analogy to the more general decomposition of a vector field into a gradient and solenoidal component \cite{GLOTZL2023127138}. The decomposition is expressed independently, and with slight variation, in the literature as the `landscape and flux framework' \cite{wang2008potential,wang2015review}, `SDE decomposition' \cite{Ao2004decomp,Yuan2017decomposition} or `GENERIC framework' \cite{duong2023generic}.\\
\\
Using the HHD, we can write the stationary probability flux and time-constant EPR in a condensed form,
\begin{align}
    \bm{J}_{ss} & = \bm{F}_{\text{irr}}P_{ss},\\
    \Phi &= \int \bm{F}_{\text{irr}}^{\top}\bm{D}^{-1}\bm{F}_{\text{irr}}P_{ss}\;d\mathbf{x},
\end{align}
where it is clear that both are driven by the presence and prevalence of irreversible drift currents.
\subsubsection{Example: the Ornstein-Uhlenbeck process}
\label{ex: OUP}
To illustrate the analysis presented in this section, we consider an analytically-tractable example in the form of the \textit{Ornstein-Uhlenbeck process} (OUP) \cite{Godreche2018OU},
\begin{align}
    \frac{d\mathbf{x}}{dt}&=-\bm{B}\mathbf{x}(t)+ \bm{\xi}(t), \label{eq: linear process}
\end{align}
where $\bm{B}$ is a constant matrix and the diffusion is spatially-homogeneous with correlations given by $\langle\bm{\xi}(t),\bm{\xi}(t')\rangle = 2\bm{D}\delta(t-t')$. This process is the stochastic extension of the simple linear differential equation, with solution,
\begin{align}
\label{eq: sol}
\mathbf{x}(t) &= e^{\bm{-B}t}\mathbf{x}(0) + \int_{0}^{t}e^{\bm{-B}(t-s)}\bm{\xi}(s)\;ds.
\end{align}
If all eigenvalues of $\bm{B}$ have positive real part, the process will converge to a unique stationary distribution defined by a zero-mean multivariate Gaussian with covariance,
\begin{align}
\label{eq: covariance}
    \bm{S}
&= 2\int_0^{\infty} e^{\bm{-B}t}\bm{D}e^{\bm{-B}^{\top}t}\;dt.
\end{align}
 It also follows from  (\ref{eq: covariance}), that the covariance satisfies the Sylvester equation \cite{Gardiner2010Methods},
\begin{align}
\label{eq: sylvester}
\bm{B}\bm{S} + \bm{S}\bm{B}^{\top}&=2\bm{D}.
\end{align}
The stationary probability flux can be written as,
\begin{align}
    \bm{J}_{ss}(\mathbf{x})&=(\bm{DS}^{-1}-\bm{B})\mathbf{x}P_{ss}(\mathbf{x}),
\end{align}
which implies that the steady-state is an ESS if $\bm{D}=\bm{BS}$ and a NESS otherwise. 
It can be further shown that the OUP is reversible if and only if $\bm{BD} = \bm{DB}^{\top}$ \cite{lax1960fluctuationsnonequilibrium}. The asymmetries of the OUP can be parametrised by the anti-symmetric matrix $\bm{Q}=\bm{BS}-\bm{D}$, which can be used to write the EPR as,
\begin{align}
    \Phi &= \int -\mathbf{x}^{\top}\bm{S}^{-1}\bm{QD}^{-1}\bm{Q}\bm{S}^{-1}\mathbf{x}P_{ss} \;\; d\mathbf{x}\\
    &=- \langle \mathbf{x}^{\top}\bm{S}^{-1}\bm{QD}^{-1}\bm{Q}\bm{S}^{-1}\mathbf{x}\rangle,
\end{align}
where $\langle \cdot \rangle = \int \cdot P_{\text{ss}}\; d\mathbf{x}$ denotes the expected value \cite{Godreche2018OU}. Finally, using that $\langle \mathbf{x}^{\top}\bm{A}\mathbf{x} \rangle = \text{Trace}(\bm{SA})$ due to the Gaussian distribution of $\mathbf{x}$, the EPR can simplified and compactly expressed as,
\begin{align}
    \Phi &= -\text{Trace}(\bm{D}^{-1}\bm{BQ}).
\end{align}
Other nonequilibrium measurements, such as the fluctuation-dissipation ratio, are also tractable for this simple process \cite{Godreche2018OU}.\\\\
Fig. \ref{fig: HHD} illustrates the HHD applied to a 2-dimensional OUP, as well as the stochastic planar limit cycle, which both have solvable NESS \cite{nartallokaluarachchi2024decomposing}. The drift field, and by extension stationary trajectories, can be decomposed into irreversible and reversible components. The reversible drift counteracts the diffusive fluctuations and maintains the process at the stationary distribution, whilst the irreversible drift drives rotation around the stationary distribution.
\subsubsection{Entropy production and irreversibility}
\label{sec: EPR irreversibility}
The relationship between the observed irreversibility of a process and its entropy production is, in general, non-trivial and unnecessary for the discussion of both brain dynamics and NESS, upon which we will focus. The interested reader is pointed towards Refs. \cite{pachter2024entropy,roldan2014thesis,Jarzynski2011equalities,seifert2012thermodynamics} for further reading. However, we present a key result bridging the irreversibility of observed trajectories and the EPR of the underlying NESS. In general, the entropy production of a stochastic trajectory has two contributions, that which is incurred during the relaxation to stationarity, and that which is produced due to probability flux in the stationary state \cite{Parrondo2009epr}. Assuming that the process relaxes to stationarity quickly, we neglect the first contribution and write the entropy production as \cite{roldan2014thesis,roldan2010dissipation},
\begin{align}
\label{eq: entropy irrev}
    \Phi = \lim_{\tau\rightarrow \infty} \frac{1}{\tau}D_{KL}[P(\{\mathbf{x}(t)\}_{t=0}^{\tau})||P(\{\mathbf{x}(\tau - t)\}_{t=0}^{\tau})].
\end{align}
Here $D_{KL}$ represents the \textit{Kullback-Leibler divergence} (KLD),
\begin{align}
    D_{KL}(P||Q) &= \int p(x)\log\left(\frac{p(x)}{q(x)}\right)\;\;dx,
\end{align}
which measures the distance between two probability distributions $P$ and $Q$ with densities $p$ and $q$ respectively \cite{thomas2006information}. In  (\ref{eq: entropy irrev}), $P(\{\mathbf{x}(t)\}_{t=0}^{\tau})$ represents the probability of a given trajectory whilst $P(\{\mathbf{x}(\tau - t)\}_{t=0}^{\tau})$ represents the probability of observing its time-reversal. Therefore the right hand side (RHS) is a natural measure of the time-irreversibility of the process. In the case that one observes partial information, in the form of coarse-grained observables, or discrete-time observations, the irreversibility becomes a lower bound on the EPR. In the case of real-world data, stochastic trajectories are often sampled at discrete time-points forming a MVTS, $\{\hat{\mathbf{x}}(t_k)\}_{k=1}^T$, then the EPR, now measured as the entropy produced \textit{per data-point}, is lower-bounded by the irreversibility of the observed MVTS \cite{roldan2014thesis,roldan2010dissipation},
\begin{align}
    \Phi &\geq \lim_{T\rightarrow \infty} \frac{1}{T} D_{KL}[P(\{\hat{\mathbf{x}}(t_k)\}_{k=1}^T)||P(\{\hat{\mathbf{x}}(t_{T-k})\}_{k=1}^T)],
\end{align}
where $P(\{\hat{\mathbf{x}}(t_k)\}_{k=1}^T)$ is the probability of observing the particular MVTS and $P(\{\hat{\mathbf{x}}(t_{T-k})\}_{k=1}^T)$ is the probability of observing its time-reversal. Here, the KLD takes the appropriate form for a distribution with discrete support $\mathcal{X}$ \cite{thomas2006information},
\begin{align}
    D_{KL}(P||Q) &= \sum_{{x}\in \mathcal{X}} p({x})\log\left(\frac{p({x})}{q(x)}\right).
\end{align}
This relationship has lead to the development of range of techniques for estimating the degree of nonequilibrium from the irreversibility of observed data \cite{roldan2010dissipation,lacasa2012irreversibility,Roldan2021irreversibility} including in neural dynamics \cite{donges2012visibility,deco2022insideout,deco2023tenet,bolton2023AoT,nartallokalu2025multilevel}.
\subsection{Discrete-space processes and Markov chains}
\label{sec: Discrete-space processes and Markov chains}
\subsubsection{Continuous-time processes}
\label{sec: continuous time}
A stochastic process defined on a finite discrete state-space describes the (time-)evolution of a sequence of random variables which take values in a discrete \textit{support}, the set of possible states. Such a process is defined by its probability distribution function, $P({X}={x},t)$, where $t$ is time, ${X}$ is the state of the process and ${x}$ is a value in the support. Without loss of generality, we consider the support to be a finite set of natural numbers, $\{1,...,n\}$. Therefore, we can consider $\bm{P}(t)$ to be the $n$-dimensional process with $P_i(t)$ being the probability of being in state $i$ at time $t$. Typically, one assumes that the process is \textit{Markovian} meaning that the dynamics only depend on the current state of the system and are \textit{memoryless}. In this case, $\bm{P}(t)$ can be described by the \textit{master equation},
\begin{align}
    \frac{d\bm{P}}{dt}&= \bm{Q}(t)\bm{P},
\end{align}
where $\bm{Q}(t)$ is the \textit{generator matrix} \cite{suhov2008markov}. Another typical assumption is that the generator matrix is constant in time, i.e. $\bm{Q}(t)\equiv \bm{Q}$, and defines a \textit{continuous-time Markov chain} (CTMC). The off diagonal entries $q_{ab}$ of $\bm{Q}$ represent the \textit{transition intensity} from state $b$ to $a$ whilst diagonal entries are defined to be $q_{aa} = -\sum_b q_{ba}$, thus each column sums to 0 which ensures that probability is conserved in time \cite{suhov2008markov}. Master equation dynamics assume that transitions occur after random, state-dependent \textit{waiting-times} which are exponentially-distributed with rate parameter $-q_{aa}$ in state $a$.\\\\
For finite state-spaces, if the process is \textit{irreducible}, i.e. each state is reachable from each other state, and \textit{aperiodic}, then it is \textit{ergodic}, meaning that it will reach a unique stationary distribution. The stationary distribution is given by the solution to $\bm{Q}\bm{P}_{ss}=0$, subject to normalisation \cite{Haggstrom2020MarkovChain}. The stationary state of a CTMC is an ESS if and only if it satisfies the (discrete) \textit{detailed balance condition} \cite{schnakenberg1976network},
\begin{align}
    q_{ab}P_{ss,b} &= q_{ba}P_{ss,a},
\end{align}
which implies that the probability flux vanishes in the steady-state and the dynamics satisfy time-reversal symmetry. To derive the EPR for a CTMC, we define the Shannon-Gibbs entropy of the discrete distribution to be,
\begin{align}
    S(t) & = - \sum_a P_a(t)\log P_a(t),
\end{align}
where upon taking the derivative, we have,
\begin{align}
    \dot{S}(t) & = -\sum_a \frac{dP_a}{dt}\log P_a(t) + \frac{dP_a}{dt},
\end{align}
where $\sum_a dP_a/dt = 0 $ by conservation of probability. Next, we use that,
\begin{align}
    \frac{dP_a}{dt} & = \sum_b (q_{ab}P_b(t) - q_{ba}P_a(t)),
\end{align}
which yields,
\begin{align}
    \dot{S}(t)& = -\sum_a\log P_a(t)\sum_b (q_{ab}P_b(t) - q_{ba}P_a(t)),\\
    &=\frac{1}{2}\sum_{a,b}(q_{ab}P_b(t) - q_{ba}P_a(t))\log \frac{P_b(t)}{P_a(t)}.
\end{align}
This expression can then be split into the heat dissipation and entropy production rate as follows,
\begin{align}
    \dot{S}(t)&=\frac{1}{2}\sum_{a,b}(q_{ab}P_b(t) - q_{ba}P_a(t))\log \frac{q_{ab}P_b(t)}{q_{ba}P_a(t)}\\
    &-\frac{1}{2}\sum_{a,b}(q_{ab}P_b(t) - q_{ba}P_a(t))\log \frac{q_{ab}}{q_{ba}}, \notag\\
    &= \Phi(t) - h_d(t).
\end{align}
Therefore, in the steady-state, we arrive at the so-called \textit{Schnakenberg formula} \cite{schnakenberg1976network},
\begin{align}
\label{eq: schnakenberg}
    \Phi = \frac{1}{2}\sum_{a,b}(q_{ab}P_{ss,b}-q_{ba}P_{ss,a})\log\left (\frac{q_{ab}P_{ss,b}}{q_{ba}P_{ss,a}}\right).
\end{align}
A CTMC can also be considered as a flow on a graph where the vertices represent discrete system states, and edges represent transition intensities. Similarly to the HHD of the stationary Langevin process, the edge-based flow defined by the thermodynamic force between system states can be decomposed into a \textit{conservative} (gradient) and non-conservative (circulating) component \cite{strang2020applications}. Specifically, by defining a flow from state $b$ to $a$ to be,
\begin{align}
    \gamma_{ab} = \log \sqrt{\frac{q_{ab}}{q_{ba}}},
\end{align}
which is analogous to the excess work required to move back and forth between these states \cite{strang2020applications, Qian2001functional}. We can decompose this edge-flow using a discrete form of the HHD,
\begin{align}
    \gamma_{ab} = (\text{curl}^{*} \varphi)_{ab} + (\text{grad} \;\beta)_{ab},
\end{align}
where $\text{curl}^{*} \varphi$ is the curl-adjoint of a flow along cycles in the graph and $(\text{grad} \;\beta)_{ab}$ is the gradient of a scalar potential defined on each vertex (see Refs. \cite{Lim2020Hodge,grady2010discretecalculus,strang2020applications} for introductions to discrete calculus and the discrete HHD). Importantly, a stationary CTMC obeys detailed balance if and only if this flow is conservative i.e. it can be written purely as the gradient of some scalar potential $\beta$,
\begin{align}
    \gamma_{ab} &= (\text{grad} \;\beta)_{ab},
\end{align}
as proved in Ref. \cite{strang2020applications}.
\subsubsection{Discrete-time processes}
\label{sec: discrete time}
Real-world data often takes the form of observations at discrete time-points. A discrete-state, discrete-time process is defined by its transition probability,
\begin{align}
    P({X}_{T+1}|\{{X}_{t}\}_{t=0}^{T}),
\end{align}
which is conditioned on the history of the process i.e. $\{{X}_{t}\}_{t=0}^{T}$. In discrete-time, the Markov condition is given by,
\begin{align}
    P({X}_{T+1}|\{{X}_{t}\}_{t=0}^{T}) = P({X}_{T+1}|{X}_{T}),
\end{align}
i.e. the transition probabilities of a \textit{discrete-time Markov chain} (DTMC) depend only on the previous state. Therefore a DTMC can be uniquely defined by its \textit{transition probability matrix} (TPM), $\bm{T}=(T_{ba})$,
\begin{align}
    T_{ba} = P({X}_{t+1}=b|{X}_{t}=a).
\end{align}
Similarly to the continuous-time case, a DTMC has a unique stationary distribution if and only if it is aperiodic and irreducible with the stationary state given by $\bm{T}\bm{P}_{ss}=\bm{P}_{ss}$, which is the Perron eigenvector of the TPM \cite{Haggstrom2020MarkovChain}. Furthermore, we can define the joint transition probability matrix (JTPM), $\bm{P}=(P_{ij})$, to be,
\begin{align}
    P_{ij} = T_{ij}P_{ss,j}.
\end{align}
The detailed balance condition can then be equivalently written as,
\begin{align}
    P_{ij}&=P_{ji},\\
    T_{ij}P_{ss,j} &= T_{ji}P_{ss,i},
\end{align}
and the EPR simplifies to,
\begin{align}
    \Phi &= \sum_{i,j}P_{ij}\log\left (\frac{P_{ij}}{P_{ji}} \right),\label{eq: KLD EPR}
\end{align}
i.e. the KLD between the forward and backward joint transition probabilities \cite{lynn2021detailedbalance}.
\section{Analysis of neuroimaging and continuous-valued recordings}
\label{sec: continuous-valued recordings}
We now proceed with the analysis of neural data, first in the form of continuous-valued neuroimaging followed by neural spike trains.\\
\\
Large-scale neural activity can be recorded using invasive and non-invasive neuroimaging techniques such as fMRI, MEG and EEG. Such data is often processed into a MVTS where each real-valued variable represents the \textit{activity} of a discrete brain region over time, subject to a specific parcellation of the brain into distinct regions \cite{deco2014greatexpectations}. The dimension of neuroimaging data can vary drastically depending on modality and experimental paradigm, with fMRI being particularly spatially precise, but temporally imprecise, and MEG being particularly spatially imprecise with greater temporal precision. Typical parcellations of the human brain can vary between $N\approx 10^0-10^6$. In high-dimensions, many traditional techniques for the analysis of stochastic dynamics such as force field reconstruction \cite{Friedrich2011complexity,Frishman2020stochasticforce} cannot be applied. Instead, one must opt for either model-based approaches, where one assumes the functional form of the dynamical system, or model-free measures which are typically correlates of the overall level of nonequilibrium.

\subsection{Model-based analysis of neural activity}
When considering model-based approaches to the analysis of nonequilibrium neural activity, one must balance two opposing forces. On the one hand, a nonlinear, biophysical model, such as a network of neural masses, may offer the rich dynamical complexity necessary to accurately model neuroimaging. However, such models may be challenging to fit from limited, noisy, and partial observations. In addition, they may not be amenable to the calculation of nonequilibrium quantities such as the EPR. On the other hand, `simple' models such as linear SDEs or discrete-state models may be easy to fit and analyse, but lack the biophysical plausibility to model the complex dynamics of the brain. In this section, we will consider a range of continuous-valued models that are used to analyse neuroimaging from a range of modalities.
\subsubsection{The linear Langevin model}
\label{sec: linear model}
\begin{figure*}
    \centering
    \includegraphics[width=\linewidth]{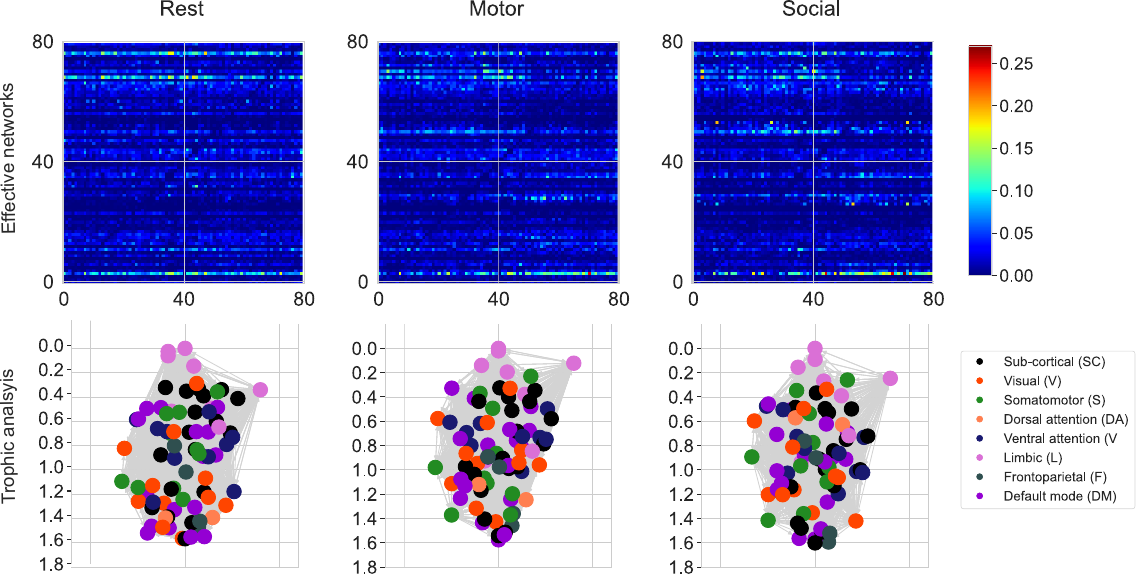}
    \caption{\textbf{Asymmetric effective networks.} Using nonequilibrium model-based approaches, one can infer an asymmetric effective network of connections between brain regions. This network can then be studied using the tools of directed networks such as \textit{trophic analysis} to discern the hierarchical organisation of brain regions and functional sub-networks. Analysis of hierarchical structure in directed networks is also closely linked to the discrete HHD defined earlier \cite{mackay2020directed,strang2022networkhhd}. Figure is adapted from results in Ref. \cite{nartallokaluarachchi2024broken}.}
    \label{fig: effective}
\end{figure*}
We start with the OUP introduced in Sec. \ref{ex: OUP}, which is also known as the \textit{linear Langevin model} and is given by,
\begin{align}
    \frac{d\mathbf{x}}{dt}&=-\bm{B}\mathbf{x}(t)+ \bm{\xi}(t),
\end{align}
where $\bm{B}$ is a constant matrix and the diffusion is given by $\langle\bm{\xi}(t),\bm{\xi}(t')\rangle = 2\bm{D}\delta(t-t')$. As the model is linear, its exact solution, HHD and nonequilibrium quantities can be obtained exactly \cite{DaCosta_2023,Godreche2018OU}. This model can be further restricted to evolve over the edges of a positively weighted network $\bm{W}$,
\begin{align}
\frac{d\mathbf{x}}{dt}&= \Theta(\gamma \bm{W}-\bm{I})\mathbf{x}(t) + \bm{\nu}(t),\label{eq: networkOU}
\end{align}
where $\bm{I}$ is the identity matrix, $\Theta$ is the reversion rate and $\gamma$ is the coupling parameter \cite{scwarze2021motifs}. For the network-based model, the thermal fluctuations are assumed to be uncorrelated between nodes, thus the additive noise satisfies,
\begin{align}
\langle \bm{\nu}(t)\bm{\nu}^{\top}(t')\rangle = 2\sigma \bm{I} \delta(t-t'),
\end{align}
where $\sigma$ is the noise intensity. This is known as \textit{isotropic diffusion}.\\\\
As mentioned earlier, a typical target for whole-brain modelling is to replicate the FC matrix of the empirical data, calculated with some pairwise similarity measure \cite{patow2024wholebrain}. In the case of the linear model, the exact covariance matrix, $\bm{S}$, of the model satisfies,
\begin{align}
\label{eq: sylvester2}
\bm{B}\bm{S} + \bm{S}\bm{B}^{\top}&=2\bm{D},
\end{align}
i.e. a Sylvester/Lyapunov equation that can be solved exactly using Kronecker algebra \cite{rotella1989kronecker} or approximately using numerical methods \cite{simoncini2016matrix}. Furthermore, we define the \textit{time-lagged covariance} to be,
\begin{align}
    \bm{S}_{\tau}& = e^{-\bm{B}\tau}\bm{S},
\end{align}
where $\tau$ is a small time lag. As the model covariance and lagged covariance can be calculated without direct sampling, they becomes a natural choice for model-fitting. Specifically, it allows for the construction of efficient gradient-descent type methods to estimate the $\bm{B}$ and $\bm{D}$ matrices that best match the empirical matrices $\bm{S}_{\text{data}}$ and $\bm{S}_{\tau,\text{data}}$ \cite{Gilson2016EC,gilson2023OU,Gilson2020EC}. In particular, the (lagged) covariance of the data can be measured as,
\begin{align}
    S_{ij, \text{data}} & = \frac{1}{T}\sum_{0\leq t\leq T}[x_i(t)-\Bar{x}_i][x_j(t)-\Bar{x}_j],\\
    S_{ij, \tau=1, \;\text{data}} & = \frac{1}{T-1}\sum_{0\leq t\leq T-1}[x_i(t)-\Bar{x}_i][x_j(t+1)-\Bar{x}_j],
\end{align}
where $\Bar{x}_i$ is the mean of the time-series $x_i(t)$ and the lag is set to $\Delta t=1$, the difference between time-points, known as the \textit{repetition time} (TR) in neuroimaging. In addition, gradient descent methods can be modified to only include connections found in an anatomical SC matrix, in an effort to maintain plausibility \cite{Gilson2016EC}.\\\\
Alternatively, linear models can be fit directly to empirical time-series using autoregressive methods \cite{nartallokaluarachchi2024broken}. First, we discretise the SDE to obtain,
\begin{align}
\mathbf{x}(t_{i+1}) &= [\bm{I}-\Delta t\bm{B}]\mathbf{x}(t_i) + \bm{\Lambda}\bm{\eta}_i,
\end{align}
where $\bm{\Lambda}\bm{\Lambda}^{\top} =2\bm{D}$ and $\bm{\eta}_i$ is an $N$-dimensional Gaussian vector with mean 0 and variance $\Delta t$. This directly mirrors the first-order autoregressive model (AR1) commonly found in statistical modelling of time-series \cite{Rencher2012multivariate},
\begin{align}
    \bm{X}(t_{i+1}) = \bm{A}\bm{X}(t_i) + \bm{\chi}(t_i),
\end{align}
where $\bm{A}$ can be calculated using ordinary least-squares autoregression and $\bm{\chi}$ is a MVTS of residuals \cite{nartallokaluarachchi2024broken}. Thus we can associate the discretised OUP to the MAR model using the following relations,
\begin{align}
\bm{A} &= \bm{I}-\Delta t\bm{B},\\
\text{Cov}[\bm{\chi}] & = 2 \Delta t \bm{D}.
\end{align}
Once $\bm{B}$ and $\bm{D}$ have been obtained, and provided that the eigenvalues of $\bm{B}$ all have positive real-part, then nonequilibrium quantities such as the EPR, fluctuation-dissipation ratio, and probability flux can be obtained from the analytical formulas as laid out in Sec. \ref{ex: OUP} or Ref. \cite{Godreche2018OU}.\\
\\
The linear model has been used to find that the EPR of human neuroimaging increases with both consciousness level \cite{gilson2023OU} and in task compared to rest \cite{nartallokaluarachchi2024broken}. In the case of the network-restricted model of Eq.~(\ref{eq: networkOU}), one can also study the asymmetric directed structure of the inferred effective network \cite{nartallokaluarachchi2024broken,Gilson2016EC,Gilson2020EC}. For example, the overall level of network asymmetry is shown to increase alongside the EPR in task compared to rest \cite{nartallokaluarachchi2024broken,benozzo2024linearstatespace}.\\\\
In addition, the hierarchy of effective networks can be unravelled using ranking algorithms from network science \cite{nartallokaluarachchi2024broken,mackay2020directed}. Fig. \ref{fig: effective} shows effective networks extracted from fMRI recordings obtained during task and at rest \cite{nartallokaluarachchi2024broken}. Using trophic analysis \cite{mackay2020directed}, a hierarchy of brain regions can be extracted and interpreted from the perspective of known functional brain networks, such as the Yeo brain networks \cite{yeo2011restingnetworks}.\\\\
The major drawback of the linear model is precisely its \textit{linearity}. Neuronal dynamics are intrinsically nonlinear \cite{Izhikevich2006dynamical}, suggesting that the linear model is a coarse and imprecise one. However, recent work has indicated that linear models may outperform nonlinear ones in the description of macroscopic resting-state dynamics \cite{nozari2024linearbrain}. Whether this is a natural limitation of current neuroimaging techniques is yet to be determined. Nevertheless, due to the computational and theoretical difficulties of fitting high-dimensional nonlinear systems to observed data, the linear model is a useful and powerful, albeit crude, technique that has shown the ability to yield valuable results.
\subsubsection{The Kuramoto model}
\label{sec: Kuramoto}
The \textit{Kuramoto model} describes interacting phase-oscillators and can be used to model synchronisation in complex systems \cite{Kuramoto1984oscillations}. Each oscillator is described by its phase, $\theta_i \in [0,2\pi]$, with collective dynamics,
\begin{align}
    \dot{\theta}_i = \omega_i + \frac{k}{N}\sum_{j=1}^N\sin(\theta_j-\theta_i),
\end{align}
where $\omega_i$ is the natural frequency, $k$ is the coupling strength and $N$ is the number of oscillators. If the coupling strength is sufficiently strong, the oscillators with align their phase and thus become \textit{synchronised}. The level of synchronisation in the system can be measured by the complex \textit{order parameter},
\begin{align}
    Re^{\bm{i}\psi}& = \frac{1}{N}\sum_{j=1}^N e^{\bm{i}\theta_j},
\end{align}
where $R\in [0,1]$ with $R=0$ at \textit{incoherence} (i.e. asynchrony) and $R=1$ at synchronisation \cite{Strogatz2000kuramoto}. The model can then be simplified,
\begin{align}
    \dot{\theta_i} = \omega_i + kR\sin(\psi-\theta_i),
\end{align}
where it is clear that oscillators are coupled through the mean-field of the system. Variations in the natural frequencies of the oscillators can be modelled by adding isotropic noise to the system,
\begin{align}
    \dot{\theta}_i = \omega_i + \frac{k}{N}\sum_{j=1}^N\sin(\theta_j-\theta_i) + \xi_i,
\end{align}
with $\langle \xi_i(t),\xi_j(t')\rangle = 2 D\delta_{ij}\delta(t-t')$, to define the \textit{Sakaguchi model} \cite{Sakaguchi1988cooperative}. This SDE can be written as a linear FP equation with $N$ spatial dimensions, as in Eq.~(\ref{eq: FP}). Alternatively, we can consider the limit $N\rightarrow \infty$ and define $P(\theta, t, \omega)$ to be the density of oscillators with phase $\theta$ and frequency $\omega$ at time $t$. Additionally, we define $g(\omega)$ to be the density of the frequencies. This density evolves according to a nonlinear\footnote{A Fokker-Planck equation is \textit{nonlinear} if the drift or diffusion directly depends on the density. In this case, the drift depends on the density through the order parameter.} FP equation \cite{Strogatz2000kuramoto},
\begin{align}
    \frac{\partial P}{\partial t} & = -\frac{\partial}{\partial \theta}(fP) +D\frac{\partial^2 P}{\partial \theta^2},\\
    f(\theta, t, \omega) & = \omega + kR\sin(\psi-\theta),
\end{align}
where the order parameter takes its continuum form,
\begin{align}
    Re^{\bm{i}\psi}& = \int_{0}^{2 \pi}e^{\bm{i}\theta}\int_{\infty}^{\infty}P(\theta,t,\omega)g(\omega)\; d\omega\;d\theta.
\end{align}
The nonlinear FP admits a trivial stationary solution, $P \equiv 1/2\pi$, which corresponds to $R=0$ and an incoherent steady-state. As the coupling strength goes to infinity, a fully synchronised steady-state with $\theta_i=\psi$ becomes possible \cite{Acebrón2005kuramoto}. More interestingly, for finite coupling a \textit{partially synchronised} steady-state emerges, where oscillators are either in the \textit{`locked'} or \textit{`drifting'} population. The density of locked oscillators follows a Dirac-delta distribution,
\begin{align}
    P(\theta,\omega) = \delta\left(\theta - \psi - \arcsin\left(\frac{\omega}{kR}\right)\right),
\end{align}
whilst drifting oscillators have the distribution,
\begin{align}
    P(\theta,\omega) =\frac{Z^{-1}}{\omega-k R\sin (\theta-\psi) },
\end{align}
where $Z$ is a normalisation constant. This steady-state bifurcates from the trivial solution at the critical coupling strength, $k=k_c$, given by \cite{Strogatz2000kuramoto},
\begin{align}
    k_c & = \frac{2}{\pi g(0)}.
\end{align}
Many extensions of the Kuramoto model have been considered with applications to both nonequilibrium statistical mechanics and neural dynamics \cite{Acebrón2005kuramoto,Rodrigues2016kuramoto,Breakspear2010kuramoto}.\\\\
Firstly, the addition of a \textit{frustration} parameter, $\alpha$, can be used to define the \textit{Sakaguchi-Kuramoto model}\footnote{These extensions to the Kuramoto model are confusingly named. We refer to the traditional model as the Kuramoto model; the model with additional noise as the Sakaguchi model; and the model with frustration as the \textit{Sakaguchi-Kuramoto} model.}
\begin{align}
    \dot{\theta}_i = \omega_i + \frac{k}{N}\sum_{j=1}^N\sin(\theta_j-\theta_i-\alpha),
\end{align}
which leads to broken symmetry in the interactions and nonequilibrium effects \cite{Gupta2014kuramoto} including the formation of \textit{turbulent waves} \cite{Kuramoto1984oscillations,Kawamura2007noiseinduced} and \textit{chimera states} \cite{Shanahan2010metastable}. Both turbulent waves and chimera states take the form of dynamic partial synchronisation, which has since been observed in neural dynamics \cite{deco2020turbulence,Deco2025turbulencereview,Hancock2025metastability}.\\
\\
Another extension of the model is to consider the addition of \textit{inertia}, which raises the dynamics to second order,
\begin{align}
    \frac{d\theta_i}{dt}&= v_i,\\
    m\frac{dv_i}{dt}&= \Tilde{k}R\sin(\psi-\theta_i),
\end{align}
where $m$ is the inertia parameter. This system is Hamiltonian and can be described by \cite{Gupta2014kuramoto},
\begin{align}
    H&=\sum_i^N \frac{mv_i^2}{2} + \frac{\Tilde{k}}{2N}\sum_{i,j} 1-\cos(\theta_i-\theta_j),
\end{align}
where it coincides with a classical thermodynamic model known as the \textit{$XY$-model} \cite{chaikin2012condensedmatter}, which becomes nonequilibrium in the presence of frustration \cite{sune2019clock}.\\
\\
In applications to neuroscience, we often consider the Kuramoto model on a complex network,
\begin{align}
    \dot{\theta}_i = \omega_i + \frac{k}{N}\sum_{j=1}^NA_{ij}\sin(\theta_j-\theta_i),
\end{align}
where $A_{ij}$ is the binary or weighted coupling from nodes $j$ to $i$ \cite{Rodrigues2016kuramoto}. As a result, it has been used to model the array of complex patterns of FC in human neuroimaging, where $\bm{A}$ is taken to be the SC of the human brain \cite{Cabral2011kuramoto}, as well as phase synchronisation in memory processes \cite{Fell2011phase}. Although the model is more complex than the linear process considered earlier, through its simple formulation it remains more amenable to analysis, including control and parameter-fitting methods \cite{Menara2022functional}, than more complex neural-mass models.
\subsubsection{The asymmetric Hopf model}
\label{sec: Hopf}
\begin{figure*}
    \centering
    \includegraphics[width=\linewidth]{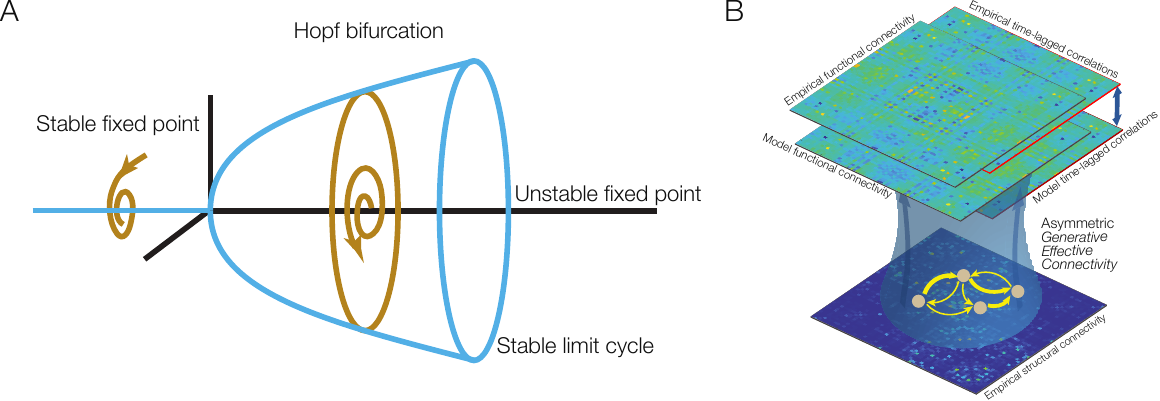}
    \caption{\textbf{Asymmetric Hopf model.} A. A Hopf bifurcation describes the qualitative change of a fixed point from stable to unstable, and the creation of a stable limit cycle. Using the bifurcation, one can define an oscillator, which is either oscillating or inactive, depending on its local parameters and network coupling. B. We begin with the symmetric structural connectivity of the human brain network and fit asymmetric connections in order to match the functional connectivity and time-lagged correlations between model and data. This yields a network of \textit{effective} connections known as the \textit{generative effective connectivity} \cite{kringelbach2024thermodynamics}. Figure is adapted from Ref. \cite{patow2024wholebrain}.}
    \label{fig: GEC}
\end{figure*}
The \textit{Hopf model} is a system composed of a network of coupled oscillators that has been used extensively to model resting and task-based neuroimaging data \cite{deco2017resting,poncealvarez2024hopf}. A \textit{Hopf bifurcation} occurs when a stable fixed point changes to an unstable fixed point and creates a limit-cycle, as illustrated in Panel A of Fig. \ref{fig: GEC} \cite{strogatz2008nonlinear}. This allows one to define a \textit{Hopf oscillator}, also known as a \textit{Stuart-Landau} oscillator \cite{Kuramoto1984oscillations}, which switches from inactive to oscillatory, depending on the value of the bifurcation parameter. Coupling such oscillators together, the Hopf model can be written in complex form as,
\begin{align}
    \frac{dz_j}{dt}&= (a_j + \bm{i}\omega_j)z_j - |z_j|^2z_j+\sum_{k=1}^NA_{jk}(z_k-z_j)+\eta_j,\label{eq: hopf complex}
\end{align}
for $j=1,...,N$, where $z_j=x_j + \bm{i}y_j$ is the complex-valued signal for node $j$, $\omega_j$ is the intrinsic regional frequency, $a_j$ is a bifurcation parameter, $\bm{A}$ is the network of interactions and $\eta_j$ is additive noise that is uncorrelated in time and between nodes. In the absence of noise and interactions, for $a_j<0$, the dynamics converge to a stable fixed point. The system goes through a bifurcation at $a_j=0$ after which the dynamics then converge to the stable limit-cycle. The model can also be written in Cartesian coordinates,
\begin{align}
    \frac{dx_j}{dt}&= (a_j  -x_j^2-y_j^2)x_j - \omega_jy_j+\sum_{k=1}^NA_{jk}(x_k-x_j)+\eta_j,\\
    \frac{dy_j}{dt}&= (a_j  -x_j^2-y_j^2)y_j + \omega_jx_j+\sum_{k=1}^NA_{jk}(y_k-y_j)+\eta_j,\label{eq: hopf cartesian}
\end{align}
for $j=1,...,N$, where, in both cases, $x_j(t)$ is considered to be the activity of region $j$ over time.\\\\
Typical applications of the model rely on an empirically obtained SC matrix, $\bm{A}$, estimated from diffusion tensor imaging \cite{deco2014greatexpectations}. However, as mentioned in Sec. \ref{sec: symmetric whole brain}, the SC matrix is undirected and thus produces equilibrium, time-reversible dynamics \cite{deco2023violations}. In order to fit both the model parameters and the effective network, heuristic approaches must be employed.\\
\\
First, the intrinsic frequencies $\omega_j$ can be estimated directly from  empirical signals, which are typically filtered into a particular range. Second, the bifurcation parameters are fixed $a_j=-0.02$, at the brink of bifurcation, as this point has been shown to closely fit resting state activity \cite{deco2017resting}. The variance of each noise component, $\sigma$, is also fixed at $0.01$. Third, in order to fit the asymmetric connections, one considers an iterative algorithm of the form,
\begin{align}
A_{ij}&=A_{ij}+\epsilon(C_{ij,\text{data}}-C_{ij,\text{model}})-\iota(C_{ij,\tau,\;\text{data}}-C_{ij,\tau,\;\text{model}}),
\end{align}
where $\epsilon,\iota$ are learning rate parameters and the initial $\bm{A}$ is the symmetric structural connectivity obtained from DTI \cite{kringelbach2023movie}. Here, $C_{ij}$ and $C_{ij,\tau}$ are the covariance and time-lagged covariance (or correlation) matrices that are measured from the data or simulations of the model. Crucially, $C_{ij,\tau}$ is asymmetric, resulting in an asymmetric effective network $\bm{A}$, as illustrated in Panel B of Fig. \ref{fig: GEC}. This inferred network is known as \textit{generative effective connectivity} (GEC) and is used to examine the reorganisation of the brain during cognitive tasks \cite{deco2023violations}, movie-watching \cite{kringelbach2023movie} and under psychedelic and anti-depressive drugs \cite{Deco2024psilodep}.\\\\
Calculating (lagged) covariance from model simulations can be computationally costly. An alternative approach is to consider a linearisation of the Hopf model \cite{poncealvarez2024hopf,deco2023violations} which yields the linear model,
\begin{align}
    \frac{d\mathbf{u}}{dt}&= \bm{J}\mathbf{u} + \eta,
\end{align}
where $\mathbf{u}=[\mathbf{x},\mathbf{y}]$ is the $2N$-dimensional state vector, and $\bm{J}$ is the $2N \times 2N$ block matrix of the form,
\begin{align}
    \bm{J}&=\begin{pmatrix}
        \bm{J}_{\mathbf{xx}} & \bm{J}_{\mathbf{xy}}\\
        \bm{J}_{\mathbf{yx}} & \bm{J}_{\mathbf{yy}}
    \end{pmatrix},
\end{align}
where each block is an $N\times N$ matrix given by,
\begin{align}
    \bm{J}_{\mathbf{xx}} &= \bm{J}_{\mathbf{yy}}=\text{diag}(\bm{a}-\bm{G})+\bm{A}\\
    \bm{J}_{\mathbf{xy}} &= -\bm{J}_{\mathbf{yx}} = \text{diag}(\bm{\omega}),
\end{align}
and $G_i=\sum_jA_{ij}$ \cite{deco2023violations}. As this model is now linear, its covariance satisfies the Lyapunov equation (\ref{eq: sylvester2}) with $\bm{B}=-\bm{J}$. Using the linear approximation, the model covariance and lagged covariance can be calculated efficiently alleviating the need for costly simulation. Although the EPR of the Hopf model cannot be calculated nor estimated effectively, previous studies have attempted to measure the degree to which the model violates the fluctuation dissipation theorem (FDT) in order to derive a model-based measure of nonequilibrium \cite{deco2023violations}.
\subsubsection{Neural field theory}
\label{sec: neural field theory}
Following the introduction of neuroimaging and the rise of \textit{connectionism}\footnote{\textit{Connectionism} refers to the paradigm in neuroscience and machine learning, by which large networks of simple units execute computations, and by extension cognition, in a distributed fashion \cite{Minsky1972perceptrons}.}, \textit{integration} has surpassed \textit{segregation} as the dominant avenue for explaining brain function \cite{friston2011FCEC}. Although segregation focuses on linking function to individual brain areas, integration attempts to explain how connected brain regions perform computation in a distributed and emergent fashion \cite{Deco2015segint}. As a result, discrete network approaches to the brain have become the dominant paradigm in the modelling and analysis of neural recordings \cite{basset2017networkneuro}.\\
\\
In contrast, recent work has shown that a continuous, geometric approach to the structure-function relationship in the brain could better explain the observed neural dynamics \cite{Pang2023geometric}. This discovery is based on an older avenue in mathematical neuroscience referred to as \textit{neural field theory} (NFT) \cite{Cook2022neuralfield,Deco2008neuralfield,Robinson2005multiscale,Coombes2005waves,Bressloff2014waves}. The primary assumption of NFT is that the cortex can be modelled as a continuous space with the level of activity as position $x$ and at time $t$ represented by a spatiotemporal function $u(x,t)$ \cite{Cook2022neuralfield,Deco2008neuralfield}. We will consider the case where $x \in \mathbb{R}$, but point the interested reader to Ref. \cite{Coombes2005waves} for consideration of higher dimensions.\\
\\
A classical model of spatiotemporal interaction is the following integro-differential equation,
\begin{align}
    \frac{1}{\alpha}\partial_tu(x,t)& = - u(x,t)+ \int_{-\infty}^{\infty}dy\;w(y)f\circ u(x-y,t), \label{eq: neural field}
\end{align}
where $f\circ u (x) = f(u(x))$ is known as the \textit{firing rate function}, typically a sigmoidal or Heaviside function, and $w(y)=w(|y|)$ is a spatial kernel that depends on the distance between positions \cite{Coombes2005waves}. This can be written in condensed form,
\begin{align}
    \mathcal{Q}(u(x,t))&= \psi(x,t),
\end{align}
where $\mathcal{Q} = [1+\alpha^{-1}\partial_tu]$ is a linear operator, and $\psi(x,t)$ is given by the spatial convolution,
\begin{align}
    \psi(x,t) &= (w \;\otimes\; f(u) )(x,t),\\
    &= \int_{-\infty}^{\infty}w(y)f(u((x-y,t))\;dy.
\end{align}
For particular choices of the spatial kernel, $w(x)$, the Fourier transform, $\mathcal{F}(w)[k]$, has a simple polynomial structure; thus, the convolution can be simplified into a PDE. For example, for the commonly used exponential coupling, $w(x)=e^{-|x|}/2$, Eq.~(\ref{eq: neural field}) simplifies to \cite{Coombes2005waves},
\begin{align}
    (1-\partial_{xx})\psi(x,t)=f\circ u(x,t).
\end{align}
A typical extension of this model is to include spatially-dependent time-delays, realised through a spatiotemporal kernel. In this case, $\psi(x,t)$ takes the more general form,
\begin{align}
\psi(x,t) = \int_{-\infty}^{\infty}\int_{-\infty}^{\infty}ds\;dy\;K(x-y,t-s)f\circ u(y,s),
\end{align}
where $K$ is a spatiotemporal kernel, for example,
\begin{align}
    K(x,t) = w(x)\delta\left(t-\frac{|x|}{\nu}\right),
\end{align}
as considered by Jirsa and Haken \cite{Jirsa1996fieldtheory}, where $\nu$ is the characteristic speed of spike propagation. By choosing $w(x)$ to be an exponential-decay of the form,
\begin{align}
    w(x)& = \frac{\exp\left(-\frac{|x|}{\gamma}\right)}{2\gamma},
\end{align}
where $\gamma$ determines the spatial rate of decay of the interactions, the model reduces to the \textit{neural wave equation}, given by the PDE \cite{Deco2008neuralfield,Jirsa1996fieldtheory,Robinson1997propogation},
\begin{align}
    \left[\frac{1}{\gamma^2}\frac{\partial}{\partial^2} + \frac{2}{\gamma}\frac{\partial }{\partial t} + 1-r^2\nabla^2 \right]u&= \left(1+\frac{1}{\gamma}\frac{\partial}{\partial t}\right)f(u),
\end{align}
where $\nabla^2$ is the Laplace operator and $\gamma = \nu/r$. This model produces travelling wave patterns that mimic the propagation of neural activity across the cortex.\\
\\
These continuous models display a range of spatially-extended phenomena such as travelling waves \cite{Coombes2005waves}, pattern-formation \cite{Amari1977patterns,Ermentrout1998patterns}, and have even led to mathematical theories of visual hallucination \cite{Ermentrout1979halluciantions,Bressloff2001hallucinations}. Several experimental studies support the existence and function of travelling waves in the brain \cite{Bolt2022waves,Raut2021waves,Mohan2024waves,Koller2024waves,Roberts2019waves} (see Ref. \cite{Muller2018travellingwaves} for a review). These include diverse behaviours such as the emergence of spiral patterns in neural activity \cite{Xu2023spiralwaves,Huang2010spiral,Townsend2015complexwaves}, a nonequilibrium phenomena that is closely linked to notions of turbulence in other areas of statistical and fluid mechanics \cite{Kuramoto1984oscillations,Eyink2006hydrodynamics}. Finally, recent work has shown that neural field models may reproduce the FC of human neuroimaging data with greater accuracy than network-based neural mass models \cite{Pang2023geometric}.
\subsubsection{Inferring stochastic models from trajectories}
\label{sec: Inferring stochastic models from trajectories}
\begin{figure*}
    \centering
    \includegraphics[width=\linewidth]{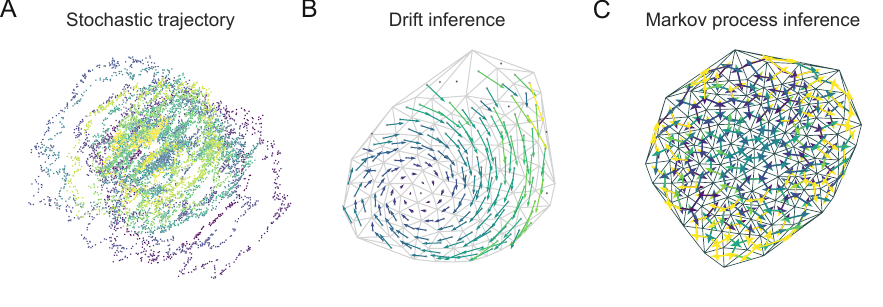}
    \caption{\textbf{Stochastic model inference.} A. An example of a stochastic trajectory in phase-space. Typically, one considers an ensemble of trajectories assumed to be from the same stationary process. B. Maximum-likelihood estimation can be used to infer the drift field of the process directly from the data and obtain a discrete approximation of a continuous model. C. Maximum-likelihood estimation can also be used to infer a discrete-state Markov process where states represent portions of state-space defined by a grid or by clustering.}
    \label{fig: model inf}
\end{figure*}

An alternative to fixed-form model-based approaches is to attempt to infer a stochastic model directly from observed trajectories \cite{Friedrich2011complexity}. Within such an approach, one considers a general stochastic dynamic such as the overdamped Langevin equation,
\begin{align}
    \dot{\mathbf{x}}(t) & = \bm{F}(\mathbf{x}) + \bm{\xi}(\mathbf{x},t), \;\;\; \mathbf{x}\in \mathbb{R}^N
\end{align}
with $\langle \bm{\xi}(\mathbf{x},t),\bm{\xi}(\mathbf{x},t')\rangle = 2\bm{D}(\mathbf{x})\delta(t-t')$. Next, given an ensemble of stochastic trajectories, one aims to infer an approximation to the drift field $\bm{F}$ and diffusion tensor $\bm{D}$. This task typically involves a discretisation of phase-space into a collection of \textit{bins}. One can then infer the drift and diffusion with an pair of estimators. A classic approach is to use the \textit{maximum-likelihood estimators} (MLE),
\begin{align}
    \hat{\bm{F}}(\mathbf{x}) &=\frac{1}{N(\mathbf{x})}\sum_{i,r\;:\; \mathbf{x}_{t_i,r}\in \mathbf{x}}\frac{\mathbf{x}_{t_{i+1},r} - \mathbf{x}_{t_{i},r}}{\tau},\\
    \hat{\bm{D}}(\mathbf{x}) &=\frac{1}{N(\mathbf{x})}\sum_{i,r\;:\; \mathbf{x}_{t_i,r}\in \mathbf{x}}\frac{[\mathbf{x}_{t_{i+1},r} - \mathbf{x}_{t_{i},r}]\otimes [\mathbf{x}_{t_{i+1},r} - \mathbf{x}_{t_{i},r}]}{2\tau},
\end{align}
where $N(\mathbf{x})$ is the number of points in each bin, $\{\mathbf{x}_{t_1},...,\mathbf{x}_{t_T}\}_{r=1}^{R}$ is an ensemble of $R$ $d-$dimensional trajectories of length $T$, $\otimes$ is the outer product and $\tau$ is the time between observed time-points \cite{Friedrich2011complexity}. Panel B of Fig. \ref{fig: model inf} shows an example of drift inference from an ensemble of stochastic trajectories from an OUP.\\
\\
A number of variations and applications of such approaches have been developed, with a particular focus on the analysis of active Brownian particles and mesoscopic biological systems \cite{garcia2018reconstruction,bruckner2024learning,battle2016brokendetailedbalance,Li2019quantifying,Gonzalez2019experimental}. Extensions include higher-order discretisations which can be employed for more accurate inference \cite{ferretti2020generallangevin}, machine-learning based approaches \cite{casert2024learning,genkin2021learning,Gao2024learning}, Bayesian inference \cite{elbeheiry2015inferencemap} or reconstruction of discrete-space flows \cite{nartallokaluarachchi2024decomposing,battle2016brokendetailedbalance}. `Stochastic force inference' \cite{Frishman2020stochasticforce}, and its formulation for the under-damped Langevin \cite{bruckner2020inferring}, is one such approach that allows for the accurate approximation of force fields in up to six dimensions without discretisation of phase-space. Additionally, it provides a natural measure of the EPR and has been applied to study the nonequilibrium dynamics of living matter \cite{Gnesotto2020learning}. These approaches, to our knowledge, have yet to be applied to neural data, but represent a promising direction in the analysis of nonequilibrium time-series.
\subsection{Discrete-state models}
\label{sec: discrete state}
Whilst neuroimaging data naturally evolves in a continuous state-space, activity can often be organised into recurrent patterns of co-activation, often referred to as \textit{brain states} \cite{kringelbach2020states,vidaurre2017hierarchically}. Examples of such persistent states include the so-called \textit{Yeo functional networks} which are patterns of FC that are often found in resting state fMRI \cite{yeo2011restingnetworks}. A description of brain activity that relies on discrete, recurrent states admits a formulation as a Markov chain, where the techniques described in Sec. \ref{sec: discrete time} can be used to study the nonequilibrium dynamics underlying the neural activity. In order to define a discrete-state model, one must first define a discrete-state space $\{1,...,n\}$. Once the continuous MVTS has been organised into a sequence of discrete states, a discrete-state model can be inferred. If the model is a DTMC, the TPM can be estimated using the MLE \cite{billingsley1961markov},
\begin{align}
    T_{ba} &= \frac{N_{a\rightarrow b}}{\sum_c N_{a\rightarrow c}},
\end{align}
where $N_{a\rightarrow b}$ is the number of observed transitions from state $a$ to state $b$. In the case of the CTMC, the transition intensities can be estimated using the MLE,
\begin{align}
    q_{ba}&= \frac{N_{a\rightarrow b}}{T_a},
\end{align}
where $T_a$ is the total holding time in state $a$ \cite{billingsley1961markov}. When the continuous-time process is observed in discrete time-steps, the MLE does not have a simple closed form, but can be estimated using expectation-maximisation algorithms \cite{Bladt2005statisticalinference}. For practical implementations, one can include a \textit{pseudo-count} for each transition, by initialising $N_{a\rightarrow b}=1$, which eliminates singularities in the EPR and corresponds to a uniform Bayesian prior for the transition probabilities \cite{manning2008laplacesmoothing}. Panel C of Fig. \ref{fig: model inf} illustrates an example of an inferred CTMC from an ensemble of stochastic trajectories from the OUP \cite{nartallokaluarachchi2024decomposing}.\\\\
The mathematical formulation and statistical inference of a discrete-state model is clearly much simpler than the continuous-space models. However, the central difficulty comes with choosing a method to classify a continuous recording into discrete states. In the following, we will describe two approaches, a naive clustering approach and the more involved Gaussian hidden Markov model. Other approaches for inferring discrete state models include deep learning approaches such as `VAMPnets' \cite{wu2020vamp,mardt2018vampnet} or alternative hidden Markov models \cite{Masaracchia2023dissecting}.
\subsubsection{Clustering approaches}
\label{sec: clustering}
When employing a clustering approach to define discrete states, the goal is to map `similar' patterns of neural activity to the same discrete state. Previous approaches have directly applied clustering to participant-concatenated time-series \cite{lynn2021detailedbalance,Cornblath2020temporal,singleton2022receptor}. Other approaches include the clustering of time-varying correlation matrices such as the so-called \textit{dynamic functional connectivity} (DFC) \cite{preti2017dynamic}. DFC is calculated by selecting a \textit{window-length}, $T_w$. Next, the FC is calculated as a matrix for each time $t_k$, as the window is moved through time,
\begin{align}
    DFC_{ij}(t_k) &= r(\{x_i(t_l)\}_{t_l=t_k}^{t_l=t_k+T_w},\{x_j(t_l)\}_{t_l=t_k}^{t_l = t_k+T_w}),
\end{align}
where $r$ is a pairwise similarity function such as PCC. The result is a matrix time-series where similar matrices can then be clustered together. An extension of this method is to employ \textit{dimensionality reduction} to the matrices prior to clustering. One such method is \textit{leading eigenvalue decomposition} (LEiDA) \cite{cabral2017leida}, an approach based on principal component analysis, which has been used to define brain states and model transitions between them \cite{kringelbach2020states,deco2019awakening}.\\\\
The most commonly used clustering algorithm is $k$-means \cite{lloyd1982kmeans} which minimises the sum of squared distances from each point in each cluster to its corresponding \textit{centroid}. Ref. \cite{saxena2017clustering} offers a review of alternative clustering algorithms. Many clustering approaches, including $k$-means, require that the number of clusters or states is pre-specified. In many cases, the appropriate number is not known and  must be chosen according to other principles, e.g., the \textit{Silhouette criterion} \cite{Kaufman1990groups} which determines the consistency of a particular clustering.
\subsubsection{The Gaussian hidden Markov model}
\label{sec: HMM}
Although clustering approaches tend to maximise the intra-cluster similarity, they treat MVTS as point clouds in high-dimensional space, thereby neglecting the temporal information present in the data. An alternative approach that accounts for the temporal propagation is to use a \textit{hidden Markov model} (HMM) \cite{eddy2004hmm}. An HMM assumes that the observed MVTS is dependent, in some known fashion, on a latent discrete-state Markov process. In fitting an HMM, one aims to simultaneously learn the discrete states and the TPM from the observed series. HMMs have found applications in the analysis of neuroimaging during rest \cite{vidaurre2017hierarchically,Vidaurre2018spontaneous}, sleep \cite{Stevner2019discovery}, memory \cite{Higgins2021replay} and cognition \cite{Baldassano2018representation}. Implementations of a number of HMMs are available is the \textit{osl-dynamics toolbox} \cite{Gohil2024osl}. In this section, we focus on the \textit{Gaussian HMM} (GHMM) \cite{vidaurre2023gaussian}.\\\\
The GHMM assumes that brain activity is a stochastic observable $\mathbf{Y}_t$ that is generated from a conditional distribution $\mathbf{Y}_t|\mathbf{X}_t$ where $\mathbf{X}_t$ is an underlying, and unknown, discrete-state Markov process. The aim of this model is to infer transition rates between these unknown states directly from the noisy observations. First, one must specify the number of states, $n$, assumed to be in the support of the HMM, $\mathbf{X}_t$. In the GHMM, each state $k$ has an associated mean, $\bm{\mu}_k$, and covariance $\bm{\Sigma}_k$. The process, at each time $t$ is thus given by,
\begin{align}
    P(\mathbf{Y}_t|\mathbf{X}_t=k) \sim \mathcal{N}(\bm{\mu}_k,\bm{\Sigma}_k),
\end{align}
i.e. at each time-point, the process is a sample from the Gaussian distribution associated with state $k$. In addition, the discrete-state of the process evolves according to a DTMC with TPM, $\bm{T}$. Extensions include the Gaussian \textit{linear} HMM (GLHMM) where the process is a dependant variable on some other independent variable $\mathbf{U}_t$,
\begin{align}
     \mathbf{X}_t \sim \mathcal{N}(\bm{\mu}_k + \mathbf{U}_t\bm{\beta}_k,\bm{\Sigma}_k),
\end{align}
and $\bm{\beta}_k$ is a vector of regression coefficients \cite{vidaurre2023gaussian}. The model can be fit using Bayesian methods such as variational \cite{jordan1999variational} or stochastic variational inference \cite{hoffman2013stochastic}. A Python implementation is available in the GLHMM package \cite{vidaurre2023gaussian}. In contrast with the clustering approaches, HMMs can be fit to maximise temporal regularity, reducing their tendency to make spurious transitions between discrete states.\\\\
The discrete states discovered by any data-driven approach can then be associated with known processes and relationships in the brain. For example, discrete states can be associated to the Yeo functional networks \cite{Cornblath2020temporal} or software such as \textit{NeuroSynth} \cite{yarkoni2011neurosynth} can be used to associate brain maps to known brain functions.\\
\\
Table \ref{tab: model based} summarises the model-based approaches to nonequilibrium neural dynamics.\\\\

\begin{table*}[h]
    \centering
    \footnotesize
    \begin{tabular}{l|cccccc}
        \hline
        \textbf{Model} & Ref. & Dynamics & Topology & Data Type \\
        \hline
        Linear Langevin & Sec. \ref{sec: linear model} & Continuous & Network & Continuous \\
        Kuramoto & Sec. \ref{sec: Kuramoto} & Continuous & Network & Phase time-series \\
        Hopf & Sec. \ref{sec: Hopf} & Continuous & Network & Continuous/Oscillatory\\
        Neural field theory & Sec. \ref{sec: neural field theory} & Continuous & Geometric mesh & Continuous\\
        General Langevin & Sec. \ref{sec: Inferring stochastic models from trajectories} & Continuous & Network & Continuous\\
        Clustered Markov chain & Sec. \ref{sec: clustering} & Discrete & - & Continuous\\
        Hidden Markov model & Sec. \ref{sec: HMM} & Discrete & - & Continuous\\
        Ising & Sec. \ref{sec: ising} & Discrete & Network & Spike-trains
        \end{tabular}
    \caption{\textbf{Model-based approaches to nonequilibrium neural dynamics.} This summarises the model-based approaches described in this paper and which mathematical framework they fall under.}
    \label{tab: model based}
\end{table*}

\subsection{Model-free measures of irreversibility}
\label{sec: model free}
\begin{figure*}
    \centering
    \includegraphics[width=\linewidth]{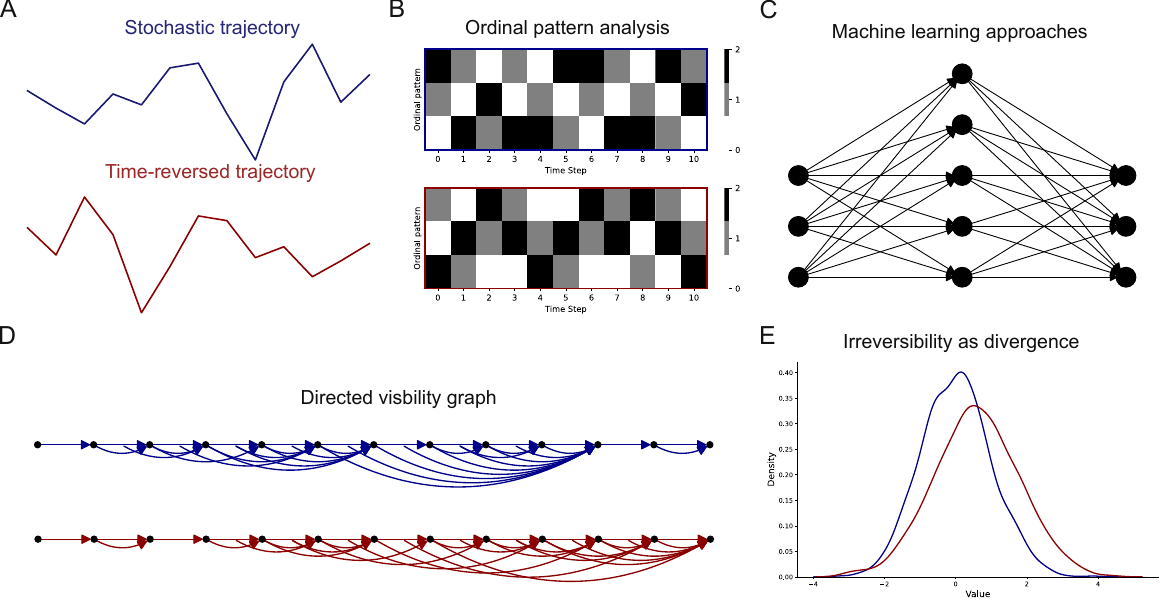}
    \caption{\textbf{Model-free measures of irreversibility.} A. Model-free measures of irreversibility typically involve a comparison between an ensemble of stochastic trajectories and their time-reversals. B. Ordinal pattern analysis is one method for converting a continuous time-series into a sequence of discrete patterns. C. Machine learning approaches can be used to classify forward and time-reversed trajectories to quantify the irreversibility of signals D. The directed visibility graph is another method for symbolisation which converts continuous time-series into a graphical structure. E. From discrete sequences or structures, the distribution of patterns, for ordinal patterns, or the directed degree distribution, for visibility graphs, can be used to quantify irreversibility through the divergence between these distributions.}
    \label{fig: symbol}
\end{figure*}
Whilst model-based approaches have been applied with some success, there remains an unavoidable trade-off between opting for a complex model with a sufficient dynamical repertoire to mimic brain dynamics, and a model that can be fit and analysed from a nonequilibrium perspective. In light of this, many methods rely on `model-free' observables that can measure the degree of nonequilibrium in time-series data, without the need of an underlying model. Many of these observables, rely on computing differences between the statistics of stochastic trajectories and their time-reversals, as illustrated in Fig. \ref{fig: symbol}.
\subsubsection{Time-lagged correlations}
Many approaches to the analysis of MVTS use \textit{auto-} and \textit{cross-correlations} (CC) to identify relationships between variables. In particular, asymmetry in CC is a fundamental hallmark of a NESS \cite{Casimir1945onsager,onsager1931reciprocal,steinberg1986noisesignals}. The `INSIDEOUT' framework \cite{deco2022insideout} is a technique that leverages this asymmetry to measure the irreversibility of neural signals. One approach is to directly measure the asymmetry between \textit{forward}, $c_f$, and \textit{backward}, $c_b$, CC,
\begin{align}
    c_{f,ij}(\tau)&=  r(x_i(t),x_j(t+\tau)),\\
    c_{b,ij}(\tau)&= r(x_i^b(t),x^b_j(t+\tau)),
\end{align}
where $\tau$ is a chosen lag time, $r$ evaluates the PCC, and $x^b$ represents a time-reversed trajectory \cite{deco2022insideout}. The level of nonequilibrium of a pair $(i,j)$ can be measured by,
\begin{align}
    I_{ij} &= |c_{f,ij}-c_{b,ij}|,
\end{align}
which depends on the choice of lag, $\tau$. Alternatively, one can consider,
\begin{align}
    TDMI_{f,ij} &= -\frac{1}{2}\log\left(1-c_{f,ij}^2\right),\\
    TDMI_{b,ij} &= -\frac{1}{2}\log\left(1-c_{b,ij}^2\right),
\end{align}
where $TDMI$ is an expression for the \textit{time-delayed mutual information} under the assumption that $x_i(t)$ and $x_j(t+\tau)$ follow a Gaussian distribution \cite{thomas2006information}. Next, the \textit{pairwise}, $I_{ij}$, and \textit{total level of nonequilibrium}, $I_T$, are given by,
\begin{align}
    I_{ij} &= |TDMI_{f,ij}-TDMI_{b,ij}|,\\
    I_T & = \sum_{ij}I_{ij}^2,
\end{align}
respectively. Using this approach, it has been shown that the level of nonequilibrium is lower in rest compared to task, in sleep compared to rest, and under anaesthetic drugs \cite{deco2022insideout}. Furthermore, there are lower levels of nonequilibrium in patients of Alzheimer's \cite{cruzat2023alzheimers} and diseases of consciousness \cite{guzman2023impairedconscious}.\\\\
CC can also be useful beyond aggregate measures of irreversibility and have been used to fit nonequilibrium models, as mentioned in Sec. \ref{sec: Hopf}, as part of the GEC approach \cite{kringelbach2024thermodynamics,kringelbach2023movie,Deco2024psilodep,deco2023violations}. In addition, CC outperform FC matrices in the classification of neural activity into tasks and states \cite{tewarie2023nonerev}.
\subsubsection{Machine learning the arrow of time}
Another framework for measuring the level of nonequilibrium in time-series data is to use a machine learning approach \cite{Seif2021machinelearning}. To do this, the problem is recast as a \textit{guessing game}. If a machine learning algorithm can correctly classify forward trajectories of a system, $\mathbf{x}_f(t)$, and backward (time-reversed) trajectories, $\mathbf{x}_b(t)$, then it has \textit{learnt} the arrow of time. More specifically, we consider a thermodynamic system defined by a Hamiltonian where $W$ is the work performed on the system during a trajectory and $F=U-TS$ is the \textit{Helmholtz free energy}, where $U$ is the internal energy, $T$ is the temperature and $S$ is the entropy \cite{Callen1985thermodynamics}. The change in free energy between the initial and final states is $\Delta F = F_B - F_A$. The probability that the trajectory was generated by the forward process is,
\begin{align}
    P(\text{forward}|\mathbf{x}_f(t))&=\frac{1}{1+\exp(-\beta(W-\Delta F))},
\end{align}
where $\beta$ is the inverse temperature of the thermal reservoir to which the system is coupled. This probability is greater than 1/2 when the change in entropy, $W-\Delta F = T(S_B-S_A) = T\Delta S$, is positive and less than 1/2 when it is negative \cite{Seif2021machinelearning}. Therefore, if a machine learning algorithm can learn the sign of $W-\Delta F$ from stochastic trajectories, then it can make optimal predictions about the direction of the arrow of time. Furthermore, the confidence or accuracy with which an algorithm predicts the direction of the arrow of time can be used as a measure of nonequilibrium as it is linked to the steady-state EPR of the process \cite{Roldán2015decision}. Both traditional techniques such as logistic regression, as well as deep learning methods such as convolutional neural networks, have been able to solve this inference problem \cite{Seif2021machinelearning}.\\\\
Inspired by this approach, the `Temporal Evolution NETwork' (TENET) framework has been developed to identify the arrow of time in brain signals \cite{deco2023tenet}. Given a collection of brain signals in a range of different conditions forming an ensemble of MVTS, one can split the available data into a training and test set. The data in both sets is then augmented to include the time-reversal of each multivariate trajectory, with the forward and backward trajectories labelled. Next, a machine-learning classifier of arbitrary design is trained using the training set to classify trajectories as forward or backward (time-reversed) examples. Next, it is applied to the test set, where the performance of the classifier can be used as a relative measure of nonequilibrium between conditions \cite{deco2023tenet}. Furthermore, this analysis can be performed independently on the time-course of individual regions or functional sub-networks to identify those with particularly nonequilibrium dynamics.
\subsubsection{Time-series symbolisation}
\label{sec: Time-series symbolisation and visibility graphs}
A central challenge for calculating information-theoretic quantities in neuroimaging, is that MVTS evolve in a continuous state-space with infinite possible values (up to the resolution of the measurement). One classic approach to solving this problem is to transform a continuous MVTS, into a univariate series of discrete \textit{symbols} \cite{Amigo2010permutation}. A very common symbolisation method was proposed by Bandt and Pompe for the calculation of \textit{permutation entropy} \cite{bandt2002permutation}, which can be applied to univariate trajectories. Given a series $x_t\in \mathbb{R}$ for $t=1,...,T$, one considers a time-delay embedding into a higher dimension $D$ \cite{kantz2003nonlinear},
\begin{align}
    \bm{y}_t=(x_t,x_{t+\tau},...,x_{t+D\tau}).
\end{align}
Each time-point in the embedded series $\bm{y}_t$ is assigned a \textit{ordinal pattern} which is a permutation of $(1,...,D)$ representing the order of the values in the embedded state-vector at that time-point. As a result, each time-point is now represented by a pattern $\bm{z}_t = \{p_{i,t}, i \in [1,...,D]\}$, where the pattern is one of $D!$ finite possible patterns or symbols. Panels A \& B of Fig. \ref{fig: symbol} illustrates ordinal pattern analysis applied to a stochastic trajectory and its time-reversal. Such a symbolisation can be used to measure the time-irreversibility of the observed trajectory \cite{martinez2018ordinal}. Given a series $x_t$ and its time reversal $\hat{x}_t$, one calculates the series of ordinal patterns for each i.e. $\bm{z}_t$ and $\hat{\bm{z}}_t$ respectively. Next, one calculates the discrete probability distributions, $P$ and $\hat{P}$, which measure the probability of observing each ordinal pattern in both the forward and backward trajectory. The KL divergence between $P$ and $\hat{P}$ becomes a natural measure of the irreversibility of the process \cite{martinez2018ordinal}.\\\\
This approach has been used to analyse the irreversibility of recordings of individual brain regions in MEG for participants with OCD and other pathologies \cite{bernardi2023ocd,zanin2020pathology}. In particular, by varying the embedding dimension $D$ or time-lag $\tau$, one can gain insight into the characteristic scale in which time-reversibility is being broken and how this differs across conditions.
\subsubsection{Visibility graphs}
An alternative approach to transforming a time-series into a discrete structure is the \textit{network analysis of time-series} \cite{varley2022nats}. This family of methods maps time-series data into a (multiplex) graph which encodes either the spatial or temporal structure of the data in its architecture. Three prominent methods include recurrence, ordinal-partition and visibility graphs. Recurrence and ordinal-partition graphs offer a natural reconstruction of a discrete-state transition graph from a univariate time-series \cite{varley2022nats}. Using the tools of nonequilibrium discrete-state processes presented in Sec. \ref{sec: Discrete-space processes and Markov chains}, one can analyse these graphs through the lens of irreversibility and the EPR.\\\\
\textit{Visibility graphs} (VG) encode temporal correlations in their degree distribution \cite{lacasa2008visibility}. Given a time-series $\{x_t\}_{t=1,...,T}$, the VG has one node for each $t=1,...,T$. Nodes $i$ and $j$ are connected by an edge if the corresponding points $(i,x_i)$ and $(j,x_j)$ are `mutually visible' which means that they satisfy that, for any intermediate data-point $(k,x_k)$, with $i<k<j$,
\begin{align}
\label{eq:visibility}
x_k<x_j+(x_i-x_j)\frac{j-k}{j-i}.
\end{align}
Geometrically, the condition is met when $(i,x_i)$ is visible from $(j,x_j)$ meaning that the line connecting $(i,x_i)$ and $(j,x_j)$ does not cross any intermediate data-points. For a MVTS, each variable is mapped into a visibility graph individually. Each graph can then be considered as a \textit{layer} within a \textit{multiplex network} as each layer contains the same nodes \cite{lacasa2015MVN}. VGs can be used to measure irreversibility by directing their edges `forward in time' i.e. if time-points $i<j$ the edge is directed $i \rightarrow j$ \cite{lacasa2012irreversibility,donges2012visibility}. The degree $d$ of each node can be decomposed into the sum of the in-going and out-going degree,
\begin{align}
    d&=d_{\text{in}}+d_{\text{out}}.
\end{align}
Panels A, D \& E of Fig. \ref{fig: symbol} illustrate an example of the directed VG constructed from a stochastic trajectory and its time-reversal. For a reversible process, it would be expected that the in- and out-degree distributions coincide; thus the divergence between these distributions is a natural measure of irreversibility. This approach has been used to study both the irreversibility and nonlinearity of EEG data in sleep \cite{Xiong2019visibility} and epilepsy \cite{donges2012visibility}. More recently, the multiplex VG has been used to measure the irreversibility of higher-order interactions in task-based MEG \cite{nartallokalu2025multilevel} as will be discussed in Sec. \ref{sec: Irreversibility across interaction-scales}.
\subsection{Beyond global irreversibility}
The majority of work analysing the nonequilibrium nature of brain dynamics has focused on quantifying the global distance from equilibrium in the form of irreversibility or the EPR. Whilst this approach has proved fruitful, highlighting the differences between a range of natural, pathological, pharmacologically-altered and task-altered brain states, the overall level of nonequilibrium remains an opaque quantity that yields limited insight into the inner workings of the brain. Advancements beyond global irreversibility include the inference of dynamics hierarchies in brain network organisation \cite{nartallokaluarachchi2024broken,Deco2024psilodep,kringelbach2024thermodynamics,kringelbach2023movie} and the use of irreversibility information to classify brain states \cite{tewarie2023nonerev}. In this section, we consider two novel applications of irreversibility analysis in continuous-space recordings of brain activity.
\subsubsection{Irreversibility and higher-order interactions}
\label{sec: Irreversibility across interaction-scales}
\begin{figure*}
    \centering
    \includegraphics[width=0.8\linewidth]{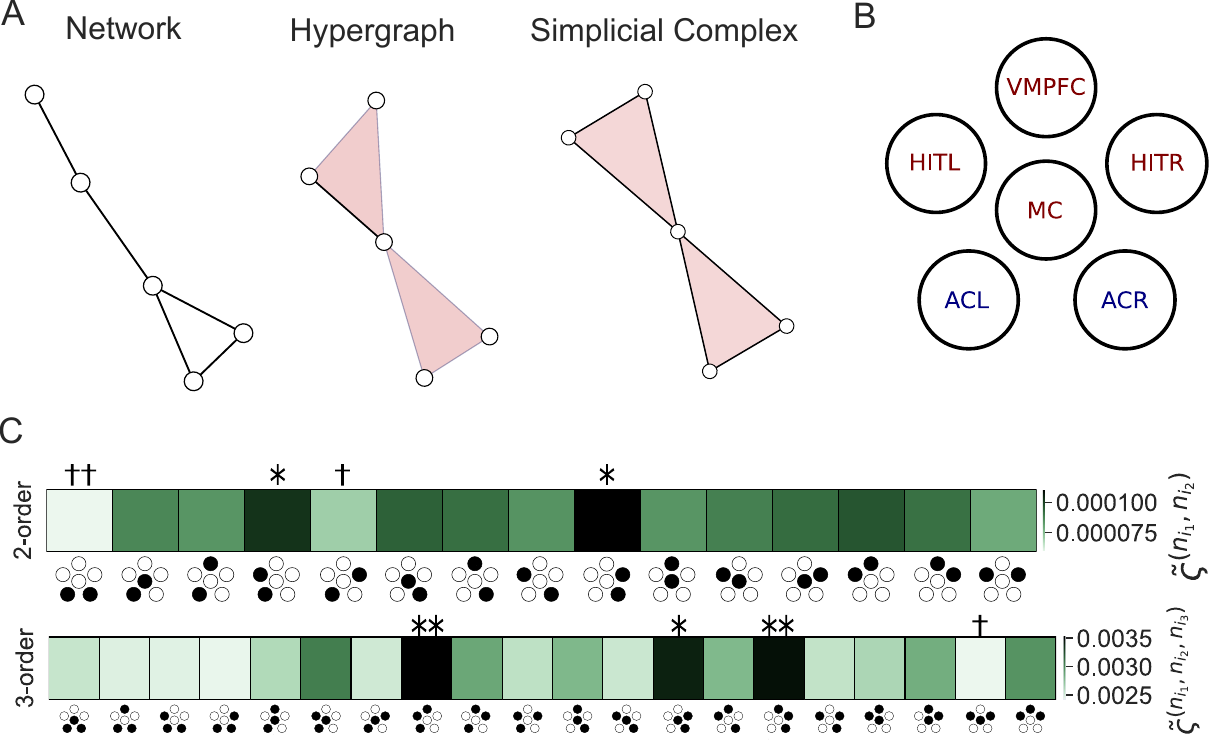}
    \caption{\textbf{Higher-order interactions from irreversible neural dynamics.} A. Higher-order interactions represent a generalisation of networks to relationships between groups of more than two variables. Higher-order interactions can either be represented by hypergraphs or simplicial complexes, where simplicial complexes necessarily require that all lower-order interactions are included to defined a higher-order interaction e.g. the existence of a triangle requires the existence of all its edges. Black lines represent pairwise edges, whilst pink triangles represent 3-order hyperedges. B. A schematic representation of 5 brain regions in the auditory system, where one aims to infer which higher-order interactions are particularly irreversible. Processing regions are denoted in red with sensory regions in blue. The regions are auditory cortices on the left and right (ACL, ACR), median cingulate gyrus (MC), ventro-medial prefrontal cortex (VMPFC) and hippocampal regions on the left and right (HITL, HITR). Adapted from Ref. \cite{nartallokalu2025multilevel} C. The irreversibility of pairwise and triplet interactions shows which pairs/triplet have a particularly irreversible interaction as measured by the DiMViGI framework. The number of (*,$\dagger$) represent the number of standard deviations the irreversibility is away from the mean across all pairs/triplets. The icon along the $x$-axis denotes each pairwise and triplet interaction in the system with black circles denoting included regions, with reference to the schematic in Panel B. Adapted from Ref. \cite{nartallokalu2025multilevel}.}
    \label{fig: higher order}
\end{figure*}
The analysis of brain network dynamics is often focused on quantifying the distributed nature of regional interactions through classical techniques like functional connectivity. However, such analysis is often restricted to pairwise interactions considering two regions in isolation. Within complex systems science, there has been renewed interest in \textit{higher-order} interactions, those that include three elements or more \cite{lambiotte2019higherorder,battinson2020physics,rosas2022behaviours,santoro2023higherorder}. As illustrated in Panel A of Fig. \ref{fig: higher order}, these constructions extend networks to structures such as simplicial complexes and hypergraphs.\\
\\
We can use the irreversibility of a multivariate trajectory to infer higher-order tuples that have a significant dynamical relationship \cite{nartallokalu2025multilevel}. More specifically, given an 
$N-$dimensional trajectory, $\mathbf{\Gamma} = \{x_1(t),...,x_N(t)\}_{t=1}^{T}$, we consider its projection into the portion of state-space defined by a particular $k$-tuple of variables $(x_{i_1},...,x_{i_k})$, to be the $k$-dimensional trajectory,
\begin{align}
    \mathbf{\Gamma}^{({i_1},...,{i_k})} = \{x_{i_1}(t),...,x_{i_k}(t)\}_{t=1}^{T},
\end{align}
where the irreversibility of $\mathbf{\Gamma}^{({i_1},...,{i_k})}$ can be measured \cite{nartallokalu2025multilevel}. Next, $k-$tuples at a particular order $k$, can be compared to find those which display a significantly higher level of irreversibility, suggesting a strongly nonequilibrium interaction between them. Although, in theory, any measure of irreversibility can be applied, a sensible approach is to measure the irreversibility in such a way that $k-$tuples can be directly compared. For example, if we were to use a model-based approach, this would involve the inference of a distinct model for each $k-$tuple. Similarly, clustering approaches will yield a unique clustering in each case. Therefore, $k-$tuples could not be compared directly. Other methods, such as cross-correlations, are naturally restricted to pairwise correlations and cannot be extended to capture higher-order dependencies.\\
\\
The multiplex visibility graphs presented in Sec. \ref{sec: Time-series symbolisation and visibility graphs} are a motivated choice as the irreversibility measurements are comparable across $k$-tuples. Using this approach, the `Directed Multiplex Visibility Graph Irreversibility' (DiMViGI) framework has been introduced and applied to investigate the higher-order organisation of interactions in the auditory cortex \cite{nartallokalu2025multilevel}. In summary, for each $k-$tuple in a MVTS, one can construct the associated directed multiplex VG, and compute the divergence between the joint in- and out- degree distribution. This yields a measure of irreversibility for each interaction in the system that can be directly compared to other tuples at a given order $k$. As a result, the analysis goes beyond a single metric for global irreversibility and yields further insight into the functional organisation of nonequilibrium brain dynamics. Panels B and C of Fig. \ref{fig: higher order} illustrate how the DiMViGI framework, applied to MEG recordings, has been used to highlight significantly irreversible pairs and triplets of regions during an auditory recognition task. In particular, it highlights the significant interaction between intra-hemispheric sensory and processing regions, as well as their synergistic interaction with medial regions.

\subsubsection{Information decomposition of irreversibility}
Another approach builds on the \textit{partial information decomposition} (PID) \cite{Williams2010nonnegative} and its variations \cite{mediano2021phiid,mediano2022pid} which have found applications in the analysis of brain network dynamics \cite{luppi2022synergy,luppi2024information}. See Refs. \cite{mediano2021phiid,mediano2022pid,Luppi2023decomposition} for further details. In summary, we consider the Markovian dynamics of two variables $\bm{x}_t = (x_t,y_t)$. The \textit{time-delayed mutual information} (TDMI) is defined as,
\begin{align}
    I(\bm{x}_t,\bm{x}_{t'}) & = D_{\text{KL}}(p(\bm{x}_t,\bm{x}_{t'})|p(\bm{x}_t)p(\bm{x}_{t'})),
\end{align}
where $p(\bm{x}_t,\bm{x}_{t'})$ is the joint distribution and $p(\bm{x}_t)$ are the marginal distributions. It can be decomposed as,
\begin{align}
    I(\bm{x}_t,\bm{x}_{t'})&= \sum_{a,b\in \mathcal{A}}I_{\partial}^{a\rightarrow b},
\end{align}
where $I_{\partial}^{a\rightarrow b}$ are \textit{information components} and $\mathcal{A}=\{R,U^{x},U^{y},S\}$ is the set of information \textit{types}, also known as \textit{atoms}. The types $R,U^{x},U^{y},S$ represent \textit{redundant, unique to $x$, unique to $y$} and \textit{synergistic} components of the information, respectively \cite{mediano2021phiid}.\\
\\
This decomposition, known as `$\Phi$ID', captures the way in which information is transformed over time within a pairwise interaction. Upon time-reversal, the total TDMI remains constant between $t$ and $t'$, yet the information dynamics are reversed. As a result, $\Phi$ID provides a family of metrics to quantify irreversibilities arising from three different \textit{kinds} of information dynamics \cite{Luppi2023decomposition}. Although the applications of such a decomposition have yet to be explored in empirical recordings of neural activity, it is known that cross-correlation measures are insufficient to capture some components of temporal irreversibility. Furthermore, such measures have been used to analyse simulated mean-field dynamics of neural populations revealing synergistic contributions to the overall level of irreversibility \cite{Luppi2023decomposition}. This approach represents a promising avenue for gaining further insight into the information dynamics underlying nonequilibrium brain dynamics.
\subsubsection{Decomposition of linear dynamics into oscillatory modes}
\label{sec: decomposition into oscillatory modes}
Neural activity is often characterised by coordinated oscillations at a range of spatial and temporal scales \cite{Buzsáki2004oscillations,Vinck2023principles}. In particular, brain dynamics can be decomposed into a collection of spatial or temporal \textit{modes} which combine to form the complex patterns that emerge in neural recordings. In particular, this has proved fruitful in the study of structure-function relationships in the brain by decomposing neural activity into contributions from spatial modes that emerge from a \textit{Laplace eigendecomposition} of structural brain connectivity \cite{Atasoy2016connectome} or geometry \cite{Pang2023geometric}.\\\\
Using such an approach, the EPR of a linear Langevin process, such as in (\ref{eq: linear process}) and Sec. \ref{sec: linear model}, was decomposed into contributions from a collection of spatial modes \cite{Sekizawa2024decomposing}. In the case of a Gaussian process with isotropic noise, the EPR of the NESS can be written as,
\begin{align}
    \sigma &= \sum_i\sum_k\sigma^{(k,i)},
\end{align}
where $\sigma^{(k,i)}$ is the contribution of the
$i$-th dimension of the $k$-th oscillatory mode with,
\begin{align}
    \sigma^{(k,i)}&= (2\pi \chi_k)^2J_k^{(i)},
\end{align}
where $\chi_k=\lambda_k/2\pi\mathbf{i}$ and $\lambda_k$ is the $k-$th eigenvalue of $-\bm{B}$, the friction matrix, appearing in Eq. (\ref{eq: linear process}). Here we have defined the following,
\begin{align}
    J_k^{(i)} = \bm{D}^{-1}_{ii}(\mathcal{F}_k\bm{S}\mathcal{F}_k^{*})_{ii},
\end{align}
where $\bm{S}$ is the covariance of (\ref{eq: sylvester}), and $\mathcal{F}_k$ is a projection matrix onto each eigenmode and $^*$ represents the matrix conjugate transpose i.e.
\begin{align}
\mathcal{F}_k=\mathbf{P}\bm{e}_k\bm{e}_k^{\top}\mathbf{P}^{-1},
\end{align}
and $\mathbf{P}$ is the matrix of eigenvectors and $\bm{e}_k$ is the $k-$th basis vector. In other words, this decomposition stems from a spectral decomposition of the friction matrix into the modes defined by its eigenbasis \cite{Sekizawa2024decomposing}.\\\\
Analysing \textit{electrocorticography} (ECoG) data in monkeys with this approach showed an increase in the EPR of anesthetized monkeys compared with resting state \cite{Sekizawa2024decomposing}, a result that disagrees with previous results showing a correlation between EPR and consciousness level \cite{guzman2023impairedconscious,gilson2023OU,sanzperl2021nonequilibrium} including a study using the same data \cite{deco2022insideout}. Nevertheless, using this approach, one is able to determine the contribution of each frequency to the overall EPR, allowing for the interpretation and integration of nonequilibrium approaches with more traditional neuroscientific results \cite{Buzsáki2004oscillations,Vinck2023principles}.

\subsection{Testing for significance, stationarity and bias in nonequilibrium measurements}
\label{sec: Testing for significance, stationarity and bias in nonequilibrium measures}
As with all analyses of empirical data, measurements of nonequilibrium quantities must be tested for statistical significance. Moreover, nonequilibrium measurements often require additional testing such as validation that the underlying data can be assumed to be sampled from a stationary process or that the observed irreversibility does not stem from the upward-bias present in finite data. In this section, we consider the methods necessary to validate the robustness and statistical validity of nonequilibrium measurements in neural data.
\subsubsection{`Bootstrapping' methods for nonequilibrium trajectories}
Some methods in the analysis of nonequilibrium dynamics are applied to an entire ensemble of stochastic trajectories or a concatenation of data between participants and scanning sessions \cite{lynn2021detailedbalance,bolton2023AoT,nartallokalu2025multilevel}. In these situations, one often obtains a single measure for a cohort of participants in a particular condition. Without a measure of uncertainty for this quantity, the statistical significance of observed differences cannot be assessed. To address this issue, we can use \textit{bootstrap methods} \cite{Horowitz2019bootstrap}. In brief, bootstrapping methods create synthetic datasets from a single one using resampling techniques. By analysing these additional bootstrapped datasets, we obtain a measure of uncertainty and error in our estimate of a numerical quantity, in this case, irreversibility. The exact implementation of a bootstrap method depends on the specific nonequilibrium measurement in question. For example, if one is considering a discrete-state model then a concatenated time-series can be written as a list of observed transitions between system states. Bootstrapped trajectories can then be generated by sampling transitions, with replacement, to generate trajectories of a given length \cite{lynn2021detailedbalance}. Similarly, if a method estimates irreversibility using an ensemble of participants or scanning sessions, e.g. DiMViGI \cite{nartallokalu2025multilevel}, then bootstrapped ensembles can be generated by sampling, with replacement, participants or scanning sessions to construct ensembles with the same number of trajectories. 
\subsubsection{Surrogate data testing and correcting for finite data}
\label{sec: finite correction}
\begin{figure*}
    \centering
    \includegraphics[width=\linewidth]{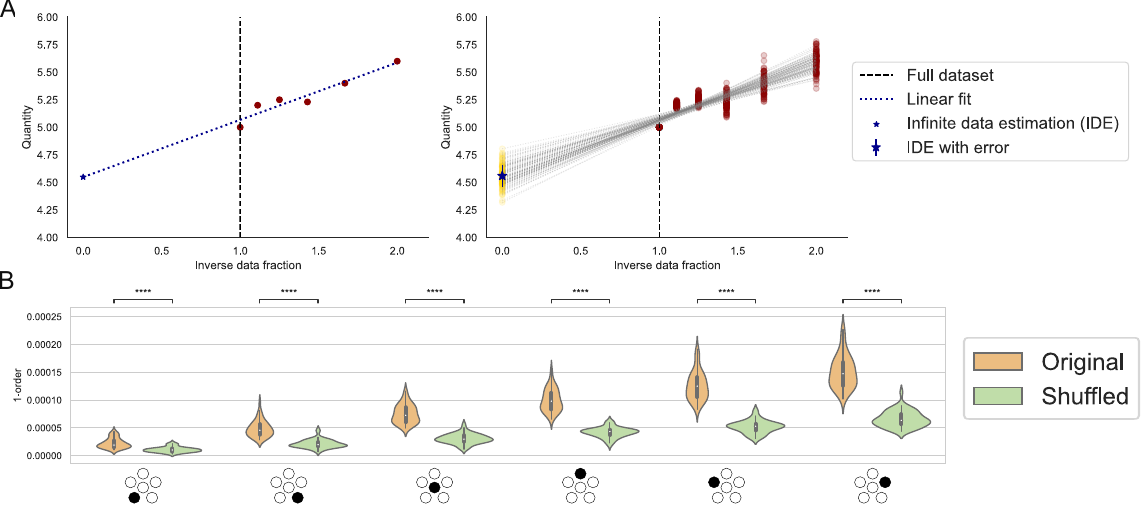}
    \caption{\textbf{Surrogate testing and correcting for finite data.} To assess the significance of measured irreversibility in neural recordings, one must consider both finite data corrections and the analysis of surrogate time-series. A. One approach to estimate quantities in the infinite data limit is to sub-sample data and measure the irreversibility. Next, one can extrapolate the relationship between the inverse data fraction (IDF) and the quantity up to IDF=0. As there are many ways of sub-sampling the data, this can be repeated to measure many different estimates of the infinite data estimate (IDE) which gives an error bar on the IDE. B. Another approach is to assess how much of the irreversibility stems from the `noise-floor', which is the irreversibility that is caused by an under-sampling of state-space. This is done by shuffling the time-series along the time-index, and then measuring the irreversibility. Here we show the irreversibility of individual regions in the auditory systems compared with surrogate shuffled data, with reference to the schematic in Panel B of Fig. \ref{fig: higher order}. We find that the irreversibility of each region is significant (****, $p<0.0001$), but also that some of the measured irreversibility is due to bias, as shuffling does not restore detailed balance perfectly. Adapted from Ref. \cite{nartallokalu2025multilevel}.}
    \label{fig: surrogate}
\end{figure*}
A similar problem in the quantification of nonequilibrium measurements from empirical data, is that there is an \textit{upward-bias} stemming from the finiteness of the data \cite{treves1995fdc}. More specifically, a finite time-series from a reversible process will have a level of irreversibility that is inherent from an under-sampling of state-space \cite{lynn2022localAoT,Roldán2015decision}. Given this constraint, how can we justify that measured irreversibility stems from genuine nonequilibrium dynamics and not from bias?\\\\
One approach is to employ \textit{surrogate data testing}, in which one generates a \textit{null time series} for comparison \cite{lancaster2018surrogate}. In order to assess the significance of irreversibility metrics, the null series must destroy the temporal correlations of the time-series thus restoring detailed balance. Any remaining irreversibility measured in the null time series is called the \textit{noise-floor} and quantifies the irreversibility resulting from an under-sampling bias \cite{battle2016brokendetailedbalance,lynn2021detailedbalance,nartallokalu2025multilevel}. Lynn et al. investigated a range of surrogate data generation methods and found that randomly shuffling time-series across the time indices restored detailed balance \cite{lynn2021detailedbalance}. More involved methods such as uni/multivariate phase-randomisation failed to restore detailed balance in the null series and thus are less suitable for such analysis. Given a distribution of nonequilibrium measurements for a collection of null series and the equivalent distribution of measurements for the original data, one can use hypothesis testing to assess if there is a significant level of irreversibility in the data. Both studies that employed surrogate-data testing found significant levels of irreversibility in neural data \cite{nartallokalu2025multilevel,lynn2021detailedbalance}, as illustrated in Panel B of Fig. \ref{fig: surrogate}.\\
\\
Although surrogate-data testing can check the significance of measured irreversibility, it does not correct for the bias in the estimates that stem from finite data. In order to correct for this bias, one must estimate the nonequilibrium measurement in the `infinite-data-limit' \cite{treves1995fdc}. One approach to this problem is to sub-sample data in a hierarchical fashion in a sequence of decreasing proportions. For example, one can calculate an information-theoretic quantity using 100\%, 80\% and 60\% of the available data. Next, we consider the relationship between the inverse data fraction and the measured quantity. By fitting a linear relationship to these points and observing the $y$-intercept, one can estimate the quantity at an inverse data fraction of zero - which corresponds to the infinite data limit, as shown in Panel A of Fig. \ref{fig: surrogate} \cite{lynn2022localAoT,treves1995fdc}. As there are many ways to sub-sample the data, this can be repeated for many random samples to produce a distribution of values in the infinite data limit and to measure the uncertainty in the corrected estimate. This approach has been used in the calculation of information theoretic quantities in neural spike trains \cite{strong1998entropy} as well as nonequilibrium measures in both spike trains \cite{lynn2022localAoT} and neuroimaging \cite{nartallokalu2025multilevel}.
\subsubsection{Testing for stationarity in neural data}
As presented in Sec. \ref{sec: math prelim}, the assumption that the distance from equilibrium can be captured in a single time-independent quantity such as the EPR, is predicated on the assumption that the data is sampled from a system in a NESS. This assumption cannot be assumed a priori, as brain dynamics may display a non-stationary behaviour \cite{Galadí2021nonstationary,Hsu2018braindynamicstate}, which can be captured by time-varying metrics such as dynamic FC \cite{preti2017dynamic}. Nevertheless, the validity of non-stationary measures in neuroimaging has been called into question \cite{Matsui2022nonstationary}, with a number of studies illustrating that brain dynamics appear to be stationary \cite{Liégeois2017temporal} and that a non-stationary behaviour in dynamic FC can be generated from stationary models \cite{Laumann2017stability,Novelli2022edgecentric}. As a result, stationarity is a common and sensible assumption, but it must be directly tested for each new dataset.\\\\
In the field of time-series analysis, particularly econometrics, a range of tests have been developed to assess the stationarity of time-series \cite{Priestly1982timeseries}. Within the context of nonequilibrium brain dynamics, the stationarity test should reflect the chosen irreversibility measurement. For example, Lynn et al. cluster neural activity into a collection of coarse-grained states \cite{lynn2021detailedbalance}. They found that the variability of the state-occupation probability did not vary significantly in time \cite{lynn2021detailedbalance}, thus justifying that the brain is operating in a NESS. A similar approach has been used to validate that neural-spike trains operate in a NESS \cite{lynn2022localAoT}. On the other hand, Nartallo-Kaluarachchi et al. fit a linear model to resting and task-based fMRI, as described in Sec. \ref{sec: linear model} \cite{nartallokaluarachchi2024broken}. As the inferred friction matrices, $\bm{B}$, satisfy the eigenvalue conditions for the model to converge to a NESS, the data can be assumed to be stationary and the EPR formula for the NESS can be applied.

\subsubsection{Preprocessing neural data}

Neural data from all modalities are contaminated by noise and artifacts. When analysing neural data, a consistent and principled preprocessing pipeline is crucial to obtain valid results. Best practices are continually evolving within the distinct communities of fMRI, MEG, EEG and electrophysiology. Here, we will briefly comment on preprocessing approaches and their possible modulation of irreversibility metrics. However, there is no correct answer on how to best preprocess data, and selecting between preprocessing options is more an art form than a hard science and left up to the neuroscientists' best judgement.\\
\\
Preprocessing stages can alter both the spatial and temporal information in a signal. For the purposes of measuring irreversibility, it is the preservation of temporal structure that is most important. For example, one may opt to use a \textit{causal filter}, which performs filtering using only previous time-steps rather than future information, implicitly respecting the arrow of time \cite{Cheveigné2019filters}. Beyond opting for causal methods when possible, we recommend following widely tested and accepted approaches such as correcting for spatial
and gradient distortions, head motion, intensity normalization, and bias field removal (in fMRI) - for example through the HCP minimal preprocessing pipeline \cite{smith2013restinghcp,glasser2013HCP}. Moreover, structural artifacts can be removed using independent component analysis approaches \cite{griffanti2014ica}. A number of open-source toolboxes exist for performing such preprocessing in a reproducible manner such as FMRIB Software Library (FSL), FreeSurfer, and FieldTrip \cite{oostenveld2011FT}.\\
\\
When analysing spike train data, such as in Sec. \ref{sec: spike train}, it is typical to have a prohibitively fine temporal granularity, thus data must be temporally coarse-grained in order to apply computational methods in a realistic time-frame. It is known that temporal coarse-graining masks irreversibility, leading to underestimates in the EPR \cite{Esposito2012coarsegraining}. Accounting for this coarse-graining can be performed via finite data correction as described in Sec. \ref{sec: finite correction}. When combining a reproducible, principled and minimalistic preprocessing pipeline with rigorous surrogate data testing, one can be more assured that results pertaining to the irreversibility of neural signals stem from genuine nonequilibrium dynamics, rather than artifacts and noise, but the possible effect of preprocessing should never be ignored.

\section{Analysis of neural spike-trains}
\label{sec: spike train}
Our analysis so far has focused on the nonequilibrium dynamics of large-scale continuous-valued recordings which are described by SDEs and represented by MVTS. At the mesoscopic scale, neuroscientists are able to capture the exact spiking behaviour of individual neurons. Due to the invasive nature of such electrophysiological recordings, they are traditionally performed in animals, with limited available data in humans. Modern high-throughput data collection techniques make it possible to record from the entire neural network of model organisms such as the \textit{C. elegans} \cite{nguyen2015celegans} and zebrafish \cite{Ahrens2013wholebrain} or from the entire visual cortex of the mouse \cite{Stringer2019visual}. Unlike the more recent interest in the nonequilibrium dynamics of neuroimaging, the dynamics of networks of spiking neurons have long been studied through the lens of nonequilibrium statistical physics. This is due to the remarkable relationship between the ubiquitous Ising model and neural spike-trains as the Ising model is both the \textit{maximum-entropy} and \textit{maximum-caliber} model for such data \cite{schneidman2006weak,presse2013maxent,bialek1996spikes,meshulam2023successes}. This analogy has opened the door to every conceivable technique associated with statistical physics models. As a result, neural spike trains have been probed for a range of thermodynamic and statistical phenomena such as criticality \cite{chialvo2010emergent,tkacik2015thermodynamics,Mora2015criticality}, crackling-noise dynamics \cite{poncealvarez2018crackling} and irreversibility \cite{lynn2022decomposing,lynn2022localAoT}.\\\\
In order to relate spiking dynamics to a discrete-state model, one starts with a \textit{raster plot}, as shown in Panel A of Fig. \ref{fig: spikes}, which records the exact spike times of each neuron in the recording. By considering a sliding window of a given length, one converts a raster plot into a discrete-time process where a neuron is \textit{active}, in a given window, if it has at least one spike, and \textit{inactive} otherwise. This coding yields a \textit{binary spike matrix}, as in Panel B of Fig. \ref{fig: spikes}, which can be considered to be an observation of a stochastic process in discrete-time with states in the space $\{0,1\}^N$, where $N$ is the number of neurons\footnote{Or $\{\pm 1\}^N$ in the case of the Ising model.}.\\\\
In this section, we first consider a model-based approach where one fits a nonequilibrium Ising model to a spike-train and then uses theoretical results to analyse the nonequilibrium properties of the data. Subsequently, we consider model-free methods, particularly a decomposition of irreversibility in spike-trains.
\subsection{The asymmetric kinetic Ising model}
\label{sec: ising}
\begin{figure*}
    \centering
    \includegraphics[width=\linewidth]{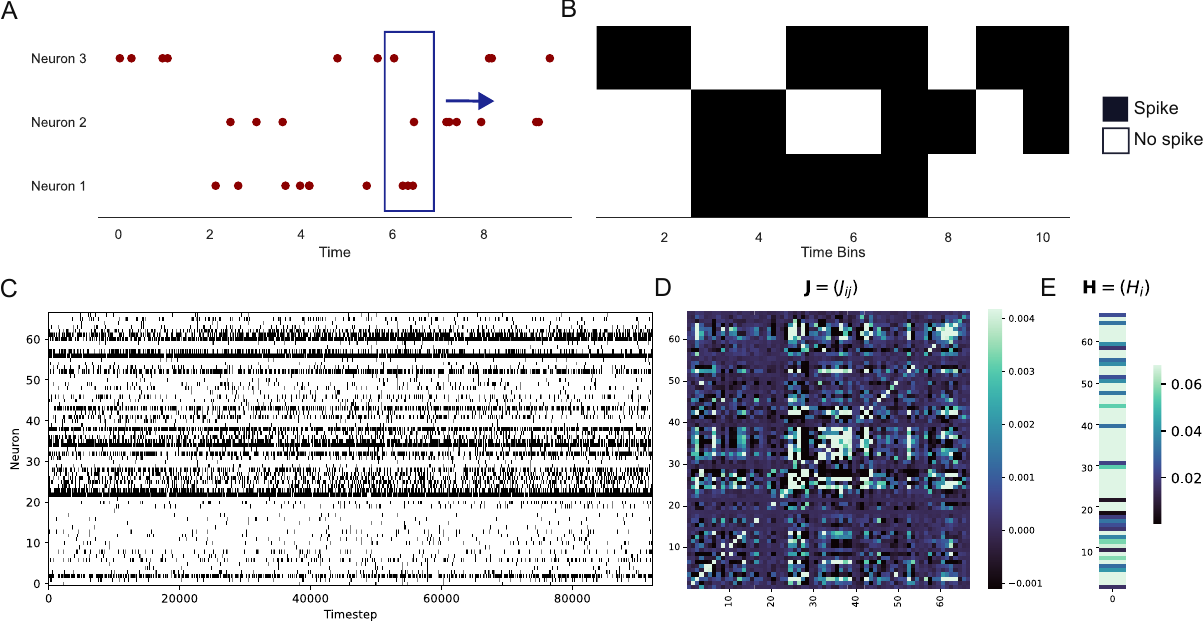}
    \caption{\textbf{Neural spike trains and the inverse Ising problem.} A. An example of a neural spike-train \textit{raster} plot. For each of three neurons, the exact times of spikes are recorded. By considering a sliding window of a certain length, this can be converted into a binary spike matrix. B. A binary spike matrix represents units that are firing in a given window with a '1' and units that are not firing with a '0'. C. A spike matrix for a population of 67 neurons. This output is comparable to the dynamics of an Ising model. D. Using gradient descent or mean-field methods, we can infer an asymmetric network of interactions $bm{J}$ where $J_{ij}$ represents the coupling strength from neuron $j$ to $i$. E. In addition, we infer the external input $H_i$ to each neuron $i$.}
    \label{fig: spikes}
\end{figure*}
The kinetic Ising model is the least-structured statistical model of binary components with delayed, pairwise interactions \cite{aguilera2021meanfield}. An Ising model contains $N$ \textit{spins} which are time-dependent variables taking values in the set $s_{i,t} \in \{\pm 1\}$. The model evolves as a DTMC with updates occurring in parallel in discrete time-steps according to the conditional transition probability,
\begin{align}
    &P(\bm{s}_t|\bm{s}_{t-1})=\prod_{i}\frac{e^{s_{i,t}h_{i,t}}}{2\cosh h_{i,t}},\\
    &h_{i,t}=H_i + \sum_jJ_{ij}s_{j,t-1},
\end{align}
where $\bm{s}_t=(s_{1,t},...,s_{N,t})$ is the state of the network at time $t$, $\bm{H}=\{H_i\}$ represents the local external fields to each spin and $\bm{J}=\{J_{ij}\}$ represents the pairwise couplings between spins \cite{aguilera2021meanfield}. When the parameters do not vary in time and the couplings are symmetric, the system obeys detailed balance and relaxes to a ESS corresponding to the system in (\ref{Eq: Boltzmann}) i.e. the maximum-entropy distribution. When the coupling are asymmetric, the dynamics are irreversible and the system relaxes to a NESS \cite{huang2014asymmetric} which is  also the maximum-caliber model \cite{presse2013maxent}.\\\\
The probability distribution of the system state is given by,
\begin{align}
    P(\bm{s}_t)&= \sum_{\bm{s}_{t-1}}P(\bm{s}_t|\bm{s}_{t-1})P(\bm{s}_{t-1}),
\end{align}
where $P(\bm{s}_{t-1})$ is the distribution at the previous time-point. The key statistical properties of the system are the \textit{mean activation rates},
\begin{align}
    m_{i,t}=\sum_{\bm{s}_t}s_{i,t}P(\bm{s}_t),
\end{align}
and the \textit{time-delayed correlations},
\begin{align}
    D_{ij,t} &=\sum_{\bm{s}_{t},\bm{s}_{t-1}}s_{i,t}s_{j,t-1}P(\bm{s}_{t},\bm{s}_{t-1})-m_{i,t}m_{j,t-1},
    \label{eq: timedelayedcorre}
\end{align}
which are sufficient statistics to define the kinetic Ising model \cite{aguilera2021meanfield}. As the Ising model is a Markov chain, its EPR is given by the Schnakenberg formula (\ref{eq: schnakenberg}). However, for this specific model, the formula can be reduced to,
\begin{align}
    \Phi_t &= \sum_{i,j}(J_{ij}-J_{ji})D_{ij,t},
\end{align}
which is time-dependent but converges to a constant value in a NESS \cite{aguilera2021meanfield}. The EPR of the Ising model is intimately related to other statistical properties. For example, both the EPR and the \textit{integrated information} are maximised at criticality \cite{martynec2020entropycriticality,aguilera2023sherrington,Aguilera2019integrated} and the EPR is driven by hierarchical asymmetry in the pairwise couplings \cite{nartallokaluarachchi2024broken}.\\
\\
In order to apply the theoretical analysis of the kinetic Ising model to the spiking behaviour of real populations of neurons, one needs to solve the so-called \textit{inverse Ising problem}. It corresponds to finding the parameters $\bm{H}$ and $\bm{J}$ that best fit observed spike-trains. Panel C of Fig. \ref{fig: spikes} illustrates an example of a raster plot for a realistic population of neurons in an electrophysiological recording. Panels D \& E illustrate an inferred coupling matrix, $\bm{J}$, and inferred external inputs, $\bm{H}$, applied to this recording. We note that the methods for fitting equilibrium and nonequilibrium or stationary and non-stationary Ising models are different \cite{hertz2013inferring}. For the asymmetric kinetic Ising model, there are two approaches that can be used to infer the parameters. First, the model can be inferred using \textit{Boltzmann learning} by maximising the log-likelihood of the model \cite{ackley1985boltzmann,hertz2013inferring,Roudi2011MF}. Second, one can perform the inference using a \textit{mean-field approximation} \cite{Roudi2011MF,hertz2013inferring,aguilera2021meanfield}. In this section, we  present both methods, but restrict our attention to the case of an asymmetric Ising model with constant parameters, as this is the most relevant to the study of NESS. Whilst Boltzmann-learning is \textit{exact}, meaning it will recover the exact parameters after infinite iterations given infinite data, it is computationally prohibitive. As a result, Boltzmann learning can be used for populations of up to approximately $N\approx100$ neurons/spins, whereas mean field approximations can be used for much larger populations.

\subsubsection{Exact inference through likelihood maximisation}
Given that the local external fields, $\bm{H}$, and the pairwise couplings, $\bm{J}$, are time-independent, we can assume that $P(\bm{s}_t)$ is stationary. Under these constraints and given a set of spike-trains, $\bm{S} = \{\bm{s}_t\}_{t=1}^T$ the \textit{log-likelihood} of the model is given by \cite{hertz2013inferring},
\begin{align}
    \mathcal{L}(\bm{S},\bm{J},\bm{H}) &= \sum_{i,t}[s_{i,t+1}h_{i,t}-\log 2 \cosh h_{i,t}].
\end{align}
This function can be ascended using the \textit{learning rules},
\begin{align}
    &\delta H_i = \eta_H [\langle s_{i,t+1}\rangle_t -\langle \tanh h_{i,t}\rangle_t ],\\
    &\delta J_{ij} = \eta_J [\langle s_{i,t+1}s_{j,t}\rangle_t - \langle \tanh h_{i,t}s_{j,t}\rangle_t],\label{eq: network update rule ising}
\end{align}
where $\eta_H, \eta_J$ are learning rates and $\langle\cdot \rangle_t$ represents a time average \cite{hertz2013inferring}. Using these learning rules, the external fields and couplings can be updated until they best match the observed spike trains (see \cite{hertz2013inferring} for details on implementation).
\subsubsection{Mean-field approximations}
A fast and effective method to fit an Ising model to an observed spike train is via a mean-field approximation (MFA). Such MFA can be systematically derived up to arbitrary order or to incorporate further temporal correlations \cite{aguilera2021meanfield}. Two very common MFA are the \textit{naive mean field} (NMF) and the more accurate \textit{Thouless-Anderson-Palmer} (TAP) approximations which have been used to fit Ising models to empirical activity from spiking neurons \cite{Donner2016Approximate,Poc-López2021inference}. Here we  focus on the NMF but refer to \cite{hertz2013inferring} for details on the TAP or algorithms for non-stationary models, and \cite{aguilera2021meanfield} for higher-order approximations.\\
\\
The NMF is given by the approximation,
\begin{align}
    m_i &= \tanh\left (H_i + \sum_j J_{ij}m_j \right ).
\end{align}
We can then use Eq. (\ref{eq: network update rule ising}), replace $s_{i,t} = m_i + \delta s_{i,t}$, and expand $\tanh$ to first order in $J_{ij}$ to obtain
\begin{align}
    \delta J_{ij} &= \eta_J[\langle \delta s_{i,t+1}\delta s_{i,t}\rangle -(1-m_i^2)\sum_kJ_{ik}\langle \delta s_{k,t} \delta s_{j,t}\rangle ].
\end{align}
As we aim to infer the exact couplings, we  assume $\delta J_{ij} = 0$ which yields the matrix equation,
\begin{align}
    \bm{D}&=\bm{AJC},
\end{align}
where $D_{ij} = \langle \delta s_{i,t+1} \delta s_{j,t}\rangle$, which is equivalent to the definition in (\ref{eq: timedelayedcorre}), $A_{ij} = (1-m_i^2)\delta_{ij}$ and $C_{ij} = \langle \delta s_{i,t} \delta s_{j,t}\rangle$ i.e. the correlation matrix. The pairwise couplings are then given by,
\begin{align}
    \bm{J} &= \bm{A}^{-1}\bm{D}\bm{C}^{-1},
\end{align}
which can then be used to infer the external fields,
\begin{align}
    H_i &= \tanh^{-1}m_i - \sum_j J_{ij}m_j.
\end{align}
\subsection{Decomposing irreversibility in spike-trains}
\begin{figure*}
    \centering
    \includegraphics[width=0.75\linewidth]{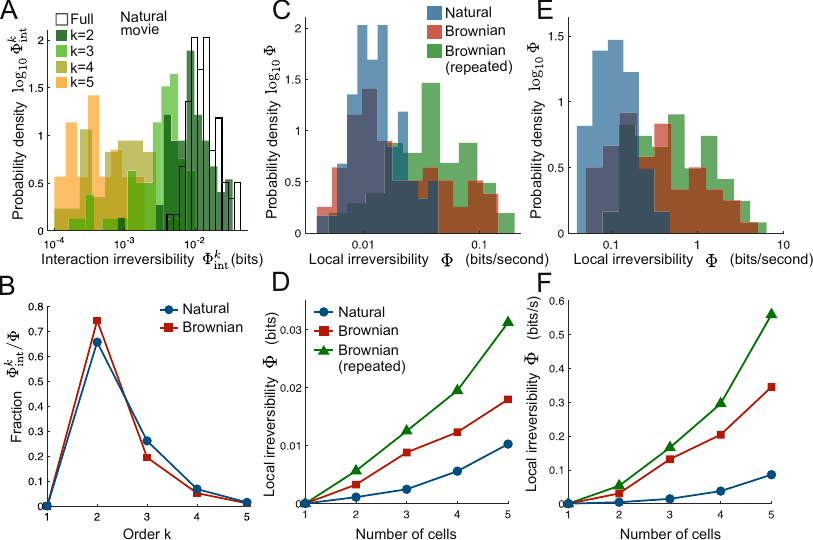}
    \caption{\textbf{Decomposing irreversibility in spike-trains}. A. Distributions of interaction irreversibility for different orders $k$ in random groups of five neurons in the retina of a salamander responding to a natural movie B. Fraction of interaction irreversibility for each order 1-5 during natural and Brownian movies C. Distribution of irreversibility for different groups of five neurons in response to natural, Brownian and repeated Brownian movies D. Local irreversibility for different stimuli. E. Same as Panel C, but normalised to account for variations in spike-rate across stimuli F. Same as Panel D, but normalised to bits/second. Adapted, with permission, from Refs. \cite{lynn2022decomposing,lynn2022localAoT}.}
    \label{fig: lynndecomposition}
\end{figure*}
As previously discussed, the arrow of time emerges due to a divergence in the probability of observing a forward and backward trajectory in a stochastic system. For Markovian systems, the EPR is given by the Schnakenberg formula (\ref{eq: schnakenberg}), which can be rewritten as,
\begin{align}
    \Phi &= \sum_{\bm{s},\bm{s}'}P(\bm{s},\bm{s}')\log \left( \frac{P(\bm{s},\bm{s}')}{P(\bm{s}',\bm{s})}\right),\label{eq: lynn EPR}
\end{align}
where $P(\bm{s},\bm{s}')$ is shorthand for $P(\bm{s}_t=\bm{s},\bm{s}_{t+1}=\bm{s}')$ which is the \textit{joint transition probability}. Eq. (\ref{eq: lynn EPR}) is a rephrasing of the EPR formula for DTMC given in Eq. (\ref{eq: KLD EPR}). If we further specialize to the case of \textit{multipartite dynamics}, where only one variable can change in a discrete time-step, then $\Phi$ can be expressed as a sum over the irreversibility of each individual variable,
\begin{align}
    &\Phi = \sum_{i=1}^N\Phi_i\\
    &\Phi_i= \sum_{\bm{s}_{-i}}\sum_{s_i,s_i'}P_i(s_i,s_i';\bm{s}_{-i})\log \left( \frac{P_i(s_i,s_i';\bm{s}_{-i})}{P_i(s_i',s_i;\bm{s}_{-i})}\right),\label{eq: multipartite EPR}
\end{align}
where $P_i(s_i,s_i';\bm{s}_{-i})$ is the probability of the $i$th variable transitioning from $s_i$ to $s_i'$ whilst the remaining variables remain fixed in the state $\bm{s}_{-i}$. Under these constraints, the EPR of a Markovian multipartite system admits two related decompositions \cite{lynn2022decomposing,lynn2022localAoT}. In the first instance, the EPR can be written as,
\begin{align}
    \Phi & = \Phi^{\text{ind}} + \Phi^{\text{int}},
\end{align}
where $\Phi^{\text{ind}}$ is the EPR assuming independent variables, given by,
\begin{align}
    \Phi^{\text{ind}} &= \sum_{i=1}^N\Phi^{\text{ind}}_i,\\
\Phi^{\text{ind}}_i&=\sum_{s_i,s_i'}P_i(s_i,s_i')\log \left( \frac{P_i(s_i,s_i')}{P_i(s_i',s_i)}\right),
\end{align}
and $\Phi^{\text{int}}$ is the remaining component of the EPR which emerges from interactions between variables \cite{lynn2022decomposing,lynn2022localAoT}.\\
\\
The component of the EPR that emerges from interactions can be further decomposed into contributions from interactions at each order. For example, the contribution from pairwise interactions can be derived by noting that,
\begin{align}
    P(s_i,s_i';s_j)&=\sum_{\bm{s}_{-i,j}}P_i(s_i,s_i';\bm{s}_{-i}),\label{eq: pairwise}
\end{align}
where $P(s_i,s_i';s_j)$ represents the joint transition probability of variable $i$ given that variable $j$ remains fixed at $s_j$. By choosing the marginal distribution, $P_i(s_i,s_i';\bm{s}_{-i})$, that minimises (\ref{eq: multipartite EPR}) and satisfies (\ref{eq: pairwise}), one calculates the minimum EPR that is consistent with observed pairwise dynamics, which is denoted $\Phi^{(2)}$. In general, one can then define $\Phi^{(k)}$, for $k=1,...,N$ where  $\Phi^{(1)} = \Phi^{\text{ind}}$ and $\Phi^{(N)} = \Phi$ \cite{lynn2022decomposing}. Finally, the unique component of the EPR that emerges from interactions at order $k$ can be defined recursively to be,
\begin{align}
    \Phi^{(k)}_{\text{int}} &= \Phi^{(k)}-\Phi^{(k-1)},
\end{align}
yielding the decomposition \cite{lynn2022decomposing,lynn2022localAoT},
\begin{align}
    \Phi &= \sum_{i=1}^N \Phi^{(i)}_{\text{int}}.
\end{align}
Crucially, this decomposition allows us to identify the scale at which irreversibility is emerging. For instance, the analysis of spike-trains from the retina of salamanders showed that the majority of irreversibility emerges from pairwise interactions between neurons, with non-trivial contributions from triplets and quadruplets of cells \cite{lynn2022decomposing}, highlighted in Panel B of Fig. \ref{fig: lynndecomposition}. Such a result is in agreement with studies showing that higher-order correlations in neural activity emerge from simple pairwise correlations between neurons \cite{schneidman2006weak,Rosch2024spontaneous}.\\\\
Another interesting conclusion from this work is that the irreversibility of the spike-activity in the retina does not coincide with the irreversibility of the stimulus, with more irreversibility emerging during natural stimulus than Brownian movies \cite{lynn2022decomposing}, illustrated in Panels A, C, D, E and F of Fig. \ref{fig: lynndecomposition}. This suggests that the irreversibility of neural activity does not reflect the stimulus - a result that mirrors the finding that human brain dynamics are more reversible during movie-watching than rest \cite{kringelbach2023movie}. For practical considerations on how to apply this decomposition to empirical data, we point the interested reader to Refs. \cite{lynn2022decomposing,lynn2022localAoT}.
\subsection{Self-organised criticality in neural activity}
\label{sec: criticality}
One of the most popular avenues for applying statistical physics techniques to neural systems is the study of \textit{criticality}. In statistical physics, criticality describes the end-point of a \textit{phase transition}, which is typically a discontinuous change in a medium in response to external conditions \cite{Stanley1987phasetransitions}. One example is the equilibrium Ising model on a 2D lattice with nearest-neighbour coupling, in the absense of external fields, where the system is operating at temperature $T$. For $T$ less than the \textit{critical} value $T_c$, the system is \textit{magnetised}, meaning the vast majority of the spins have the same state, either $\pm1$, also known as \textit{ferromagnetic long-range order}. For  temperatures higher than $T_c$, the system is globally \textit{disordered}, meaning there are no clusters of magnetised spins. At the critical point $T=T_c$, the \textit{correlation length}, which measures the spatial distance over which two spins are correlated, diverges to infinity \cite{Stanley1987phasetransitions}. The correlation length is the length scale associated with the exponential decay of the correlation function:
\begin{align}
    C(r)& = \langle s_is_j\rangle -\langle s_i\rangle \langle s_j \rangle,
\end{align}
where $r=d(i,j)$ is the distance between spins $i$ and $j$ and $\langle \cdot \rangle$ is the average. Above the critical temperature, this function decays exponentially,
\begin{align}
    C(r)\sim e^{-r/\xi},
\end{align}
where $\xi$ is the \textit{correlation length}. As $T\to T_c$, the \textit{critical point}, the correlation length diverges, $ \xi\rightarrow\infty$. This results in a \textit{self-similar} structure where clusters of magnetised states exists within other clusters, a property often referred to as \textit{scale-invariance}. Such phenomena can be examined through the framework of the \textit{renormalisation group} \cite{Wilson1983renormalisation}. Criticality can be identified from the behaviour of observables which behave as \textit{power-laws} \cite{Newman2005powerlaws},
\begin{align}
    P(T)\propto (T-T_c)^{-\alpha},
\end{align}
for some \textit{critical exponent} $\alpha$.\\
\\
Up to this point, notions of criticality focused on equilibrium systems which were controlled via modulation of an external parameter i.e. temperature. In 1987, Bak et al.,  using sand-piles as a toy model,  introduced \textit{self-organised criticality} (SOC), which proposed that dynamical systems may be attracted to the critical point without the need for external modulation \cite{Bak1987selforganised}. This notion generalised notions of criticality to nonequilibrium dynamical systems \cite{Muñoz2018criticality}. Since, then evidence of SOC have been found in an array of disparate fields, but most notably biological \cite{Mora2011biologicalcriticality,Muñoz2018criticality} and neural systems \cite{Plenz2021soc,Cocchi2017criticality}.\\
\\
In the context of neural dynamics, much of the focus has been on the emergence of \textit{avalanches} \cite{Beggs2008criticality,Beggs2003avalanches}. In order to assess if the brain is operating at, or near, the critical point, the cascade of neural activity that follows a neuron firing can be described as an avalanche whose size\footnote{This includes the duration and the number of neurons involved} can be estimated from data. If the size of avalanches in the brain follows a power-law distribution, it can be seen as evidence of SOC \cite{Beggs2008criticality}, as avalanches in critical sand-pile models follow a similar distribution \cite{Bak1987selforganised}. The \textit{critical brain hypothesis} argues that self-organising to criticality can optimise the speed of information processing in the brain due to high responsiveness to external stimulus without the risk of signals failing to propagate, nor triggering runaway excitation \cite{Beggs2008criticality}. Moreover, some systems admit a \textit{Griffith's phase}, which is an entire region of critical-like dynamics \cite{Griffiths1969nonanalytic}, as opposed to a single point. Griffith's phases have since been shown to emerge in neuronal dynamics \cite{Moretti2013griffiths}.\\
\\
Assessing the evidence for criticality in the brain remains particularly challenging \cite{Beggs2012critical}. First, confirming power-laws from empirical data is a difficult task, as  many systems can appear to be power-law in a small window of variable space when in fact they are not when considered in wider regions, and many simple methods for fitting power-laws are not robust techniques for parameter estimation \cite{Clauset2009powerlaw}. Moreover, neural recordings typically suffer from a \textit{`subsampling problem'}, where experimental observations make up a fraction of the total system, and where multiple corrective techniques must be applied to minimise biases \cite{Levina2022subsampling}. Finally, whilst nonequilibrium dynamics are not necessary for the existence of critical dynamics, they appear related. In fact, for the nonequilibrium \textit{Sherrington-Kirkpatrick} model,\footnote{This is an infinite Ising model where coupling $J_{ij}\sim \mathcal{N}(0,\Delta J)$ \cite{aguilera2023sherrington}.} it has been shown that the EPR peaks at the critical transition \cite{aguilera2023sherrington}.
\section{Nonequilibrium neural computation}
\label{sec: nonequilibrium neural computation}
Whilst we have focused on the modelling and analysis of nonequilibrium dynamics in empirical neural data, the theoretical study of computational cognitive systems operating away far from equilibrium is an emerging field \cite{yan2013landscape,DaCosta2021bayesian,Parr2019markov,friston2021parcels,kolchinsky2018semantic,Wolpert2019thermodynamics,Wolpert2024thermodynamics}. Much of this work deals with computational systems that are more abstract than the brain. Hence, here we will only discuss a single area of this adjacent field which is close to neuroscience.
\subsection{Bayesian mechanics and the `free-energy principle'}
\begin{figure*}
    \centering
    \includegraphics[width=0.75\linewidth]{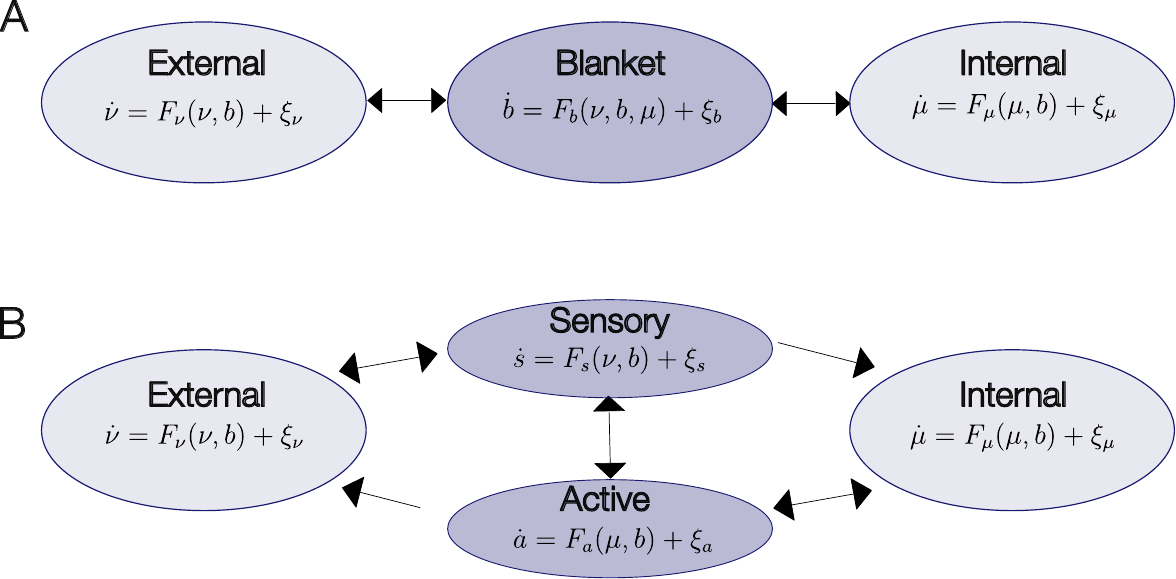}
    \caption{\textbf{Markov blankets}. A. Conditional dependence structure of external, blanket and input states in a Markov blanket. B. Blanket states can be decomposed into active and sensory states where active depend on internal states and sensory depend on external states.}
    \label{fig: markovblanket}
\end{figure*}
The \textit{`free energy principle'} (FEP) is a prominent theory of self-organisation in biology stemming from neuroscience \cite{Friston2010fep}. From the perspective of the FEP and the `Bayesian brain' hypothesis, cognition in the brain occurs via a form of Bayesian inference \cite{Friston2010fep,Knill2004bayesian}. To avoid confusion, we note that  the term `free-energy' while motivated by analogies with physical systems, is not a physical free energy but a more general concept.  More specifically, the brain aims to infer a generative internal model from sensory information coming from its environment. The separation between the internal model and the external environment is mediated through a \textit{Markov blanket} \cite{Pearl1998graphical}. A Markov blanket is comprised of three states, the internal, blanket, and external states, as shown in Panel A of Fig. \ref{fig: markovblanket}. Using visual processing as an example, the position of an object in an environment is an external state, the sensory information falling on the retina is a blanket state and the neural activity in the visual cortex is an internal state \cite{Parr2019markov}. The Markov blanket condition dictates that, given the blanket state, there is no conditional dependence between the internal and external states. Mathematically, this can be expressed as,
\begin{align}
    \nu \perp \mu | b  \Leftrightarrow p(\nu,\mu|b)=p(\nu|b)p(\mu|b), 
\end{align}
where $\nu, b$ and $\mu$ represent the external, blanket and internal states of the system respectively and $\perp$ represents conditional independence \cite{Parr2019markov}. Given the blanket state, the most likely internal and external states are defined to be,
\begin{align}
    \bm{\nu}(b) &= \mathbb{E}[\nu|b],\\
    \bm{\mu}(b) &= \mathbb{E}[\mu|b].
\end{align}
According to the FEP, cognition is performed by attempting to encode the most probable external state in the internal state via the \textit{synchronisation map},
\begin{align}
    \sigma(\bm{\mu})&=\text{argmax} \;p(\nu|b),
\end{align}
which associates internal states to the most likely external causes given sensory information \cite{DaCosta2021bayesian}. Additionally, we can associate each internal state to a probability distribution over the external states $q_{\mu}(\nu)$. Cognition through \textit{variational Bayesian inference} hinges on the maximisation of a lower-bound on the likelihood \cite{Beal2003variational} which is equivalent to minimising the `free-energy functional',
\begin{align}
    \mathcal{F}(b,\mu)\geq \mathcal{F}(b,\bm{\mu}),
\end{align}
where
\begin{align}
    \mathcal{F}(b,\mu)=D_{\text{KL}}\left [ q_{\mu} || p(\nu | b)\right] - \log p(b,\mu).
\end{align}
As states vary over time, they can be described a stochastic process $\mathbf{x}(t) = (\nu(t),b(t),\mu(t))$, which can be modelled by a Langevin equation in a NESS \cite{Parr2019markov,DaCosta2021bayesian,friston2021parcels}. Given a Langevin process, such as that considered in Sec. \ref{sec: continuous space},
\begin{align}
    &\dot{\mathbf{x}} = \bm{F}(\mathbf{x}) + \xi(t),\\
    &2\bm{D}\delta(t-t')=\langle\xi(t),\xi(t') \rangle, 
\end{align}
the condition for stationarity can be written as,
\begin{align}\label{eq: parr HHD}
    &\bm{F}=(\bm{Q}-\bm{D})\nabla (-\log P_{ss}),\\
    &\nabla \cdot (\bm{Q}\nabla P_{ss})=0,
\end{align}
for some matrix $\bm{Q}$ which controls irreversible flow. This is an alternative formulation of the Helmholtz-Hodge decomposition (HHD) presented in Sec. \ref{sec: HHD langevin} \cite{Friston2011valueattractors,Parr2019markov}. We then partition blanket states into \textit{active} and \textit{sensory} states, $b(t) = (a(t),s(t))$, where active states influence external variables given internal variables, whilst sensory states influence internal variables given external states, as shown in Panel B of Fig. \ref{fig: markovblanket} \cite{DaCosta2021bayesian,Parr2019markov}. Using the conditions of Markov blanket and the HHD, the stationary dynamics of the most likely internal states can be derived to obtain,
\begin{align}
    &\dot{\bm{\mu}}(b) = -\bm{D}_{\sigma(\mu),\sigma(\mu)}\nabla_{\bm{\mu}}\mathcal{F}(\bm{\mu},b),\\
    &\dot{\bm{a}}(b) = (\bm{Q}_{aa}-\bm{D}_{aa})\nabla_{\bm{a}}\mathcal{F}(\bm{\mu},b).
\end{align}
As a result, a Markov blanketed system operating in a NESS yields dynamics where internal and active states minimise the free energy of the blanket states (see \cite{Parr2019markov,DaCosta2021bayesian,friston2021parcels} for  details).  This result has been interpreted in both neuroscience and biology as a theory of self-organisation based on \textit{self-evidencing} and \textit{homeostasis} from a NESS \cite{Hohwy2014selfevidencing,Friston2010fep,Parr2022activeinference}.

\section{Discussion and outlook}
Approaches integrating nonlinear dynamics, statistical physics, and complex systems have generated many new innovative methods for analysing neural system recordings \cite{Turkheimer2022complexsystems,lynn2019review,basset2017networkneuro,Bassett2011complexity}. Although a relatively recent perspective in neuroscience, the study of nonequilibrium dynamics in biological systems has been essential for understanding their mechanisms and energetics \cite{fang2019nonequilibrium,Gnesotto2018brokendetailedbalance}, particularly through its connections to stochastic thermodynamics \cite{roldan2024thermodynamicprobes}. However, interpretation of nonequilibrium dynamics from the perspective of thermodynamics is less obvious for the brain since the relationship between meso/macroscopic recorded signals and microscopic energetic processes becomes blurred across scales. Nevertheless, the formalism of stochastic nonequilibrium dynamics has repeatedly proved fruitful in the analysis of neural recordings. For instance, accumulating evidence increasingly suggests that neural dynamics can be well approximated by a stochastic process in a nonequilibrium steady state \cite{lynn2021detailedbalance,nartallokaluarachchi2024broken,lynn2022localAoT} and that the entropy production rate is a robust neural correlate of consciousness \cite{guzman2023impairedconscious,delafuente2023irreversibility,gilson2023OU}. Unlike traditional neural correlates of consciousness \cite{Koch2016neural} which originate from neuroscience, nonequilibrium measures come equipped with the formal mathematical apparatus developed for stochastic thermodynamics, which allows for different, but complementary, formulations of the same phenomena. An example of this is the nonequilibrium theories of discrete Markov chains and continuous Langevin processes, presented in Sec. \ref{sec: math prelim}, both of which have been applied to the study of nonequilibrium brain dynamics from two inherently different perspectives, yet they yield consistent results.\\
\\
Nevertheless, it is crucial that this area of research goes beyond differences in nonequilibrium measures in experimental conditions. The recognition that nonequilibrium dynamics are an intrinsic feature of neural activity has become, and should remain, a key consideration in the development of new modelling and analysis tools in neuroscience. For example, as demonstrated in Sec. \ref{sec: Equilibrium models of neural activity}, undirected network models at both the single-unit and population scales produce equilibrium, time-reversible trajectories, failing to account for the nonequilibrium dynamics inherent in the data. This brings to the forefront techniques for studying asymmetries in functional, structural and effective connectivity. This can be done with a range of approaches such as  studying directed animal connectomes \cite{kale2018directed}, using directed functional measures \cite{Friston2013granger,cliff2023pairwise}, or using model-based approaches \cite{Seguin2019directionality,Tanner2024asymmetric,nartallokaluarachchi2024broken,Gilson2020EC,kringelbach2024thermodynamics}. Further, the nonequilibrium theory of Markov chains \cite{schnakenberg1976network} yields a new perspective for the analysis of traditional discrete-state models such as HMMs or clustered DFC. Such states can be associated to known networks of functional connectomes and their dynamics can be decomposed using the Helmholtz-Hodge decomposition to reveal cycles and gradient flows that govern their temporal organisation \cite{strang2020applications,schnakenberg1976network}.\\
\\
 To go beyond aggregate measures of irreversibility,  recent developments proposed to decompose nonequilibrium measurements into differential contributions from interaction orders \cite{lynn2022decomposing,lynn2022localAoT} and oscillatory modes \cite{Sekizawa2024decomposing} or  to isolate the nonequilibrium dynamics of informational components \cite{Luppi2023decomposition} or higher-order interactions \cite{nartallokalu2025multilevel}. These techniques use nonequilibrium statistical physics to form a basis for the development of new methods that provide insight into the coordination of distributed activity across neural systems and offer an alternative to traditional approaches in network neuroscience.\\
\\
The analysis of nonequilibrium dynamics in neural spike-trains represents a deviation from the typical maximum-entropy approach \cite{schneidman2006weak,meshulam2023successes}. The study of nonequilibrium spiking dynamics has led to the development of novel techniques for fitting asymmetric Ising models to experimental data \cite{aguilera2021meanfield} and has inspired theoretical advancements addressing large-scale networks operating far from equilibrium \cite{aguilera2023sherrington}. More generally, this development motivates a departure from maximum-entropy assumptions toward maximum-caliber models where nonequilibrium dynamics are front and center \cite{pachter2024entropy,presse2013maxent}. Furthermore, nonequilibrium models of neuronal activity offer the opportunity to discover relationships between nonequilibrium dynamics and other complex statistical phenomena such as criticality, which can be more closely related to neural computation \cite{Beggs2008criticality}.\\
\\
Unlike the typical noisy, low-dimensional data considered in nonequilibrium thermodynamics \cite{seifert2019inference}, neural recordings are often high-dimensional time-series of continuous or binary variables. As a result, analysis of nonequilibrium brain dynamics has called for the development of novel time-series analysis methods that scale to the dimension of neural recordings. Examples such as using time-lagged correlations \cite{deco2022insideout}, auto-regressive models \cite{nartallokaluarachchi2024broken,bolton2023AoT,gilson2023OU} or machine learning \cite{deco2023tenet} are techniques that can be applied efficiently to time-series with large dimension and have been developed and applied specifically due to the challenges and opportunities of neural data.\\
\\
A relatively unexplored area of profound importance to  nonequilibrium brain dynamics, is its behaviour across scales. More specifically, nonequilibrium dynamics can be found in the sub-cellular mechanics of individual neurons, as well as the macroscopic dynamics of coarse-grained neural activity. Coarse-graining of variables and states can have a significant and complex effect on thermodynamic quantities \cite{Egolf2000equilibrium,Esposito2012coarsegraining}, yet the interaction between these scales in the brain remains unknown. Developments in this area will provide greater insight into the metabolism and energetics of neural systems \cite{Harris2012synaptic}, and how this is reflected in neural dynamics. One approach to gain insight into this area is to consider multi-scale recordings such a simultaneous spiking and local-field potentials \cite{Buzsáki2012lfp} or simultaneous EEG-fMRI \cite{Ritter2006eegfmri}, where complimentary signals can be studied in parallel. This will bring new insight to the cascade of nonequilibrium dynamics across scales, as well as identify subtle differences in the dynamics captured by different recording modalities.\\
\\
A central question in neural dynamics, that has not yet been discussed here, is how they perform distributed computation. Notably, it is believed that computation occurs at the neuronal level, thus it is more opaque to methods focusing on large-scale brain dynamics. However, as nonequilibrium dynamics emerge across scales, their role in \textit{computation} may be significant. For example, as mentioned previously, Hopfield's model of associative memory relies on symmetric coupling and reversible dynamics to encode memories in a potential landscape \cite{hopfield1982hopfield,hopfield1984continuous}. Nonequilibrium currents, enforced via asymmetric coupling \cite{yan2013landscape}, could be used to design models of associative memory with improved capacity \cite{Behera2023active} or to induce cycling between a sequence of memories. Whilst this was explored in the wake of Hopfield's original papers \cite{Sompolinsky1986temporal, Crisanti1987asymmetric, Hertz2006irreversible}, the focus was on random asymmetric couplings, leading to limited biological relevance and interest. Modern research on nonequilibrium neural systems could rejuvenate this dormant area, leading to nonequilibrium models of associative memory.\\
\\
More modern research on computation via neural dynamics focuses on the emergence of \textit{neural manifolds} which are low-dimensional manifold structures present in high-dimensional neural recordings \cite{Langdon2023neuralmanifolds}. Analysing dynamics over the manifold \cite{Duncker2021dynamicsmanifold}, as well as symmetry breaking \cite{Pillai2017symmetry}, from the perspective of nonequilibrium dynamics could bring new insight into how neural systems compute through active dynamics. The relationship between neural systems and artificial recurrent neural networks also provides an avenue to study machine learning systems with nonequilibrium dynamics. Somewhat aside from neuroscience, applications of stochastic thermodynamics have emerged in deep learning such as through the study of stochastic optimisation algorithms \cite{Adhikari2023machinelearning} or through new architectures like diffusion models \cite{Sohl2015diffusion}.\\
\\
Nonequilibrium brain dynamics is the latest attempt to leverage tools from the quiver of dynamical systems and statistical physics for the analysis of neural dynamics across time- and length-scales. It has lead to the development of important correlates of cognitive exertion and consciousness that offer a bridge between dynamics and neuro-psychology. Beyond merely quantifying the global irreversibility of neural signals, this approach has driven the development of innovative modelling and analysis techniques that emphasize asymmetry and hierarchy in brain organization.\\
\\
Whilst theoretical, mathematical frameworks have lead to significant developments in physics, they have so far failed to answer questions of equal fundamentality in neuroscience and biology. By introducing the techniques of nonequilibrium statistical physics, we make progress towards a more complete theoretical understanding of the brain within a paradigm that holds the potential to draw links between the brain as a dynamical system, biological organ and information processing machine. Nevertheless, this field is in its infancy, and more work is needed to develop this area into a unified framework that is theoretically consistent and empirically useful.

\section*{Author contributions}
R.N-K, R.L., A.G. designed the review. R.N-K wrote the manuscript. M.L.K, G.D., R.L., A.G. edited the manuscript.
\section*{Acknowledgements}
The authors would like to thank Christopher W. Lynn for permission to adapt Fig. \ref{fig: lynndecomposition}.\\
\\
R.N.K was supported by an Engineering Physical Sciences Research Council (EPSRC) Doctoral Scholarship from Grants No. EP/T517811/1 and No. EP/R513295/1 and an Enrichment Community Award from The Alan Turing Institute. M.L.K. was supported by the Centre for Eudaimonia and Human Flourishing (funded by the Pettit and Carlsberg Foundations) and Center for Music
in the Brain (funded by the Danish National Research Foundation, DNRF117). G.D. was supported by grant PID2022-136216NB-I00 funded by MICIU/AEI/
10.13039/501100011033; by ERDF A way of making Europe, ERDF, EU, Project Neurological Mechanisms of Injury, and Sleep-like cellular dynamics
(NEMESIS; ref. 101071900), funded by the EU ERC Synergy Horizon Europe; and AGAUR research support grant (2021 SGR 00917) funded by the Department of Research and Universities of the Generalitat of Catalunya. R.L. was supported by the EPSRC grants EP/V013068/1, EP/V03474X/1 and EP/Y028872/1.
\addcontentsline{toc}{section}{References}
\bibliographystyle{IEEEtran}
\bibliography{tau}

\end{document}